\documentclass[sigconf, table]{acmart}

\AtBeginDocument{%
  }

\copyrightyear{2025}
\acmYear{2025}
\setcopyright{acmlicensed}
\acmConference[SIGIR '25] {Proceedings of the 48th International ACM SIGIR Conference on Research and Development in Information Retrieval}{ July 13--18, 2025}{Padua, Italy.}
\acmBooktitle{Proceedings of the 48th International ACM SIGIR Conference on Research and Development in Information Retrieval (SIGIR '25), July 13--18, 2025, Padua, Italy}
\acmISBN{979-8-4007-1592-1/25/07}
\acmDOI{10.1145/3726302.3729937}
\newcommand{\modelname}{CORONA }
\usepackage{algorithm}
\usepackage{algorithmic}
\usepackage[most]{tcolorbox}
\usepackage{multirow} 
\usepackage{makecell}
\usepackage{xcolor}%%
\usepackage{subcaption}
\usepackage{graphicx}
\usepackage{balance}
\usepackage{xcolor}

\author{Junze Chen}
\affiliation{%
  \institution{Beijing University of Posts and Telecommunications}
  \city{Beijing}
  \country{China}
}
\email{junze@bupt.edu.cn}

\author{Xinjie Yang}
\affiliation{%
  \institution{Beijing University of Posts and Telecommunications}
  \city{Beijing}
  \country{China}
}
\email{yangxinjie@bupt.edu.cn}

\author{Cheng Yang}
\authornote{Corresponding author.}
\affiliation{%
  \institution{Beijing University of Posts and Telecommunications}
  \city{Beijing}
  \country{China}
}
\email{yangcheng@bupt.edu.cn}

\author{Junfei Bao}
\affiliation{%
  \institution{Beijing University of Posts and Telecommunications}
  \city{Beijing}
  \country{China}
}
\email{bupt_baojunfei@bupt.edu.cn}

\author{Zeyuan Guo}
\affiliation{%
  \institution{Beijing University of Posts and Telecommunications}
  \city{Beijing}
  \country{China}
}
\email{1154459434@bupt.edu.cn}

\author{Yawen Li}
\affiliation{%
  \institution{Beijing University of Posts and Telecommunications}
  \city{Beijing}
  \country{China}
}
\email{warmly0716@126.com}

\author{Chuan Shi}
\affiliation{%
  \institution{Beijing University of Posts and Telecommunications}
  \city{Beijing}
  \country{China}
}
\email{shichuan@bupt.edu.cn}
\renewcommand{\shortauthors}{Junze Chen et al.}
\begin{document}

%\title{Chain of Retrieval on Graphs: A Framework for Graph-based Recommendation with Large Language Models}

\title{CORONA: A Coarse-to-Fine Framework for Graph-based Recommendation with Large Language Models}

%\title{CORONA: Coarse-to-Fine Graph-based Recommendation with Large Language Model Reasoning}


\begin{abstract}
Recommender systems (RSs) are designed to retrieve candidate items a user might be interested in from a large pool, with a typical approach being the use of graph neural networks (GNNs) to capture high-order interaction relationships. As large language models (LLMs) have demonstrated remarkable success across various domains, researchers are exploring ways to apply their capabilities for improving recommendation performance. However, existing work limits the use of LLMs to either re-ranking recommendation results of traditional RSs or pre-processing the datasets as data augmenters. Both lines of work failed to explore LLMs' capabilities during the filtering process of candidate items, which may lead to suboptimal performance. Instead, we propose to leverage LLMs' reasoning abilities during the candidate filtering process, and introduce \textit{Chain Of Retrieval ON grAphs (CORONA)} to progressively narrow down the range of candidate items on interaction graphs with the help of LLMs: (1) First, LLM performs preference reasoning based on user profiles, with the response serving as a query to extract relevant users and items from the interaction graph as \textit{preference-assisted retrieval}; (2) Then, using the information retrieved in the previous step along with the purchase history of target user, LLM conducts intent reasoning to help refine an even smaller interaction subgraph as \textit{intent-assisted retrieval}; (3) Finally, we employ a GNN to capture high-order collaborative filtering information from the extracted subgraph, performing \textit{GNN-enhanced retrieval} to generate the final recommendation results. The proposed framework leverages the reasoning capabilities of LLMs during the retrieval process, while seamlessly integrating GNNs to enhance overall recommendation performance. Extensive experiments on various datasets and settings demonstrate that our proposed CORONA achieves state-of-the-art (SOTA) performance with an 18.6\% relative improvement in recall and an 18.4\% relative improvement in NDCG on average. Our code is available on GitHub at https://github.com/BUPT-GAMMA/CORONA.

%that \textcolor{red}{apply \textit{LLMs after candidate filtering} treat them as mere predictors, requiring interaction and candidate set information to be incorporated into the input prompts. This design restricts LLMs to the narrow task of re-ranking recommendation results generated by traditional RSs.} Another line of work attempts to \textcolor{red}{apply \textit{LLMs before candidate filtering} treating them as augmenters to alleviate data sparsity. However these methods fail to leverage LLMs' reasoning abilities in either the filtering or re-ranking process} 

%To better utilize the rich and general knowledge possessed by LLMs, we propose to leverage LLMs \textcolor{red}{during the candidate filtering itself, rather than relying on them for fine-grained selection from a given candidates.} Therefore, we propose \textit{Chain Of Retrieval ON grAphs (CORONA)} to progressively narrow down the range of candidate items on interaction graphs: 

%Their performance is heavily constrained by the context length limitation and significant computational overhead of LLMs. We directly apply the reasoning abilities of LLMs to the retrieval process:
\end{abstract}

% \iffalse

%    <concept>
%        <concept_id>10002951.10003317.10003347.10003350</concept_id>
%        <concept_desc>Information systems~Recommender systems</concept_desc>
%        <concept_significance>500</concept_significance>
%        </concept>
%  </ccs2012>

% \ccsdesc[500]{Information systems~Recommender systems}
% \fi


\maketitle

\section{Introduction}
%p1 一句话解释rec，点出要从很多的item candicate里面去排序检索.为了捕捉大量用户/item之间的关系，一种xx的方案 graph-based...介绍graph based rec
Recommender systems aim to filter out candidate items that users might purchase~\cite{das2007google, davidson2010youtube, xu2022rethinking}. For effective predictions, it is crucial to capture collaborative filtering (CF) relationships from massive user-item interactions. To this end, graph-based approaches~\cite{wang2019neural, he2020lightgcn, yang2024graphpro, wei2020graph} typically construct a bipartite graph with historical user-item interactions, and employ graph neural networks to model high-order relationships, showing satisfied recommendation performance~\cite{gao2023survey, ying2018graph}. %推荐综述；图推荐综述

\begin{figure}[ht]
    \centering
    \begin{subfigure}[b]{\linewidth}
        \centering
        \includegraphics[width=\linewidth]{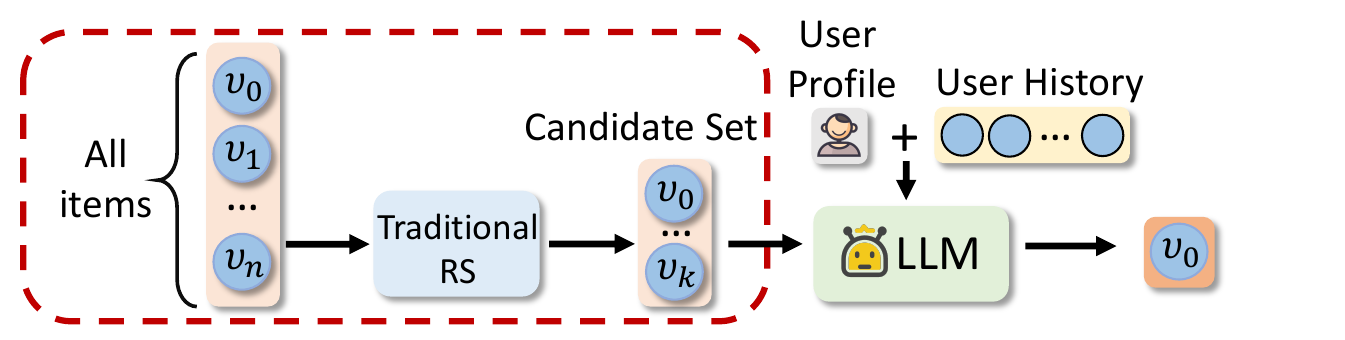}
        \caption{{Previous methods that apply LLMs after candidate filtering. These methods typically rely on traditional RSs to obtain candidate items, and use LLMs as the final predictors (\textit{e.g.,} TALLRec~\cite{bao2023tallrec}).}}
        \label{subfig1}
    \end{subfigure}
    
    \vspace{1em} % 可调整子图之间的间距
    
    \begin{subfigure}[b]{\linewidth}
        \centering
        \includegraphics[width=\linewidth]{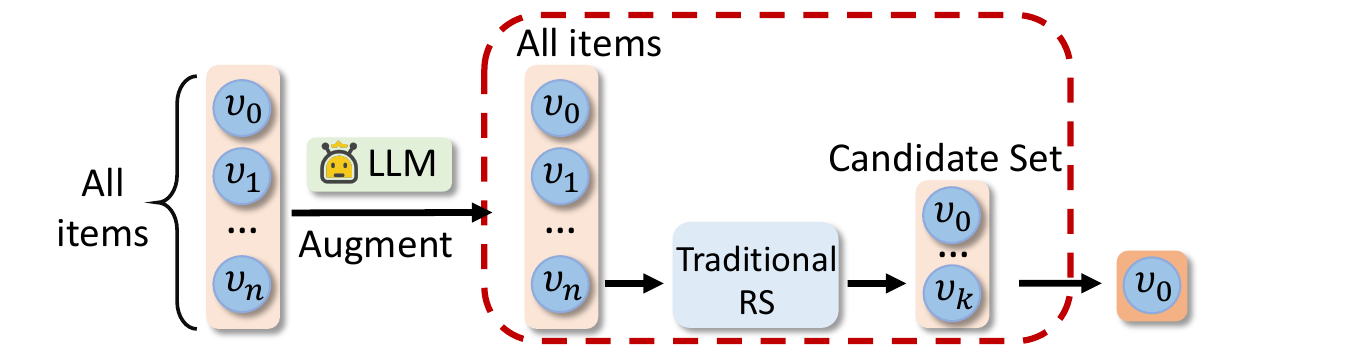}
        \caption{{Previous methods that apply LLMs before candidate filtering. These methods typically use LLMs for data augmentation, and then rely on traditional RSs to obtain candidate items (\textit{e.g.,} LLMRec~\cite{wei2024llmrec}).}}
        \label{subfig2}
    \end{subfigure}
    
    \vspace{1em} % 可调整子图之间的间距
    
    \begin{subfigure}[b]{\linewidth}
        \centering
        \includegraphics[width=\linewidth]{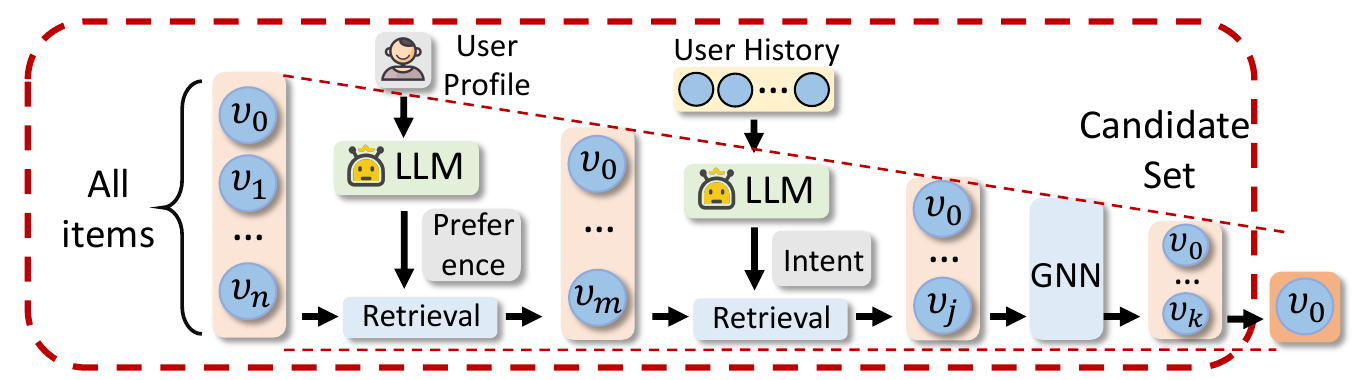}
        \caption{Our proposed framework using LLMs for coarse-grained retrieval and GNNs for fine-grained recommendation, {directly leveraging LLMs to assist candidate filtering.}}
        \label{subfig3}
    \end{subfigure}

    \caption{Comparisons between previous methods and our proposed coarse-to-fine framework, with the candidate filtering process highlighted by the red dashed box. We leverage LLMs for coarse-grained preference and intent reasoning instead of data augmentation or fine-grained item selection. Our framework progressively narrows the range of candidate items and integrates GNNs for improved performance.}
    \label{intro_fig}
    \vspace{-0.4cm}
\end{figure}

The rapid advancements of LLMs have showcased remarkable capabilities in generating, reasoning, and modeling world knowledge. Recommender systems are also anticipated to reap significant benefits from the development of LLMs. Recent efforts have leveraged LLMs for recommendation and achieved promising performance across various scenarios~\cite{liao2024llara, hou2024large, bao2023tallrec, zhang2023recommendation, gao2023chat, hou2024large, wang2023rethinking, kim2024large, zhu2024collaborative, ren2024enhancing, zhang2024large, wang2024user, dai2023uncovering, zhao2024aligning}. Existing approaches can be categorized into two types: 
(1) \textit{Applying LLMs after candidate filtering}. These methods typically integrate relevant information (\textit{e.g.,} item candidates) into natural language-based prompts, and derive recommendations from LLM-generated responses. Due to the context length limitations of LLMs, these methods need traditional RSs to generate candidate item sets as inputs. Early attempts in this category relied on in-context learning (ICL) with frozen LLMs~\cite{gao2023chat, hou2024large}, while later methods began to explore different fine-tuning strategies for specialization~\cite{bao2023tallrec, zhang2023collm, geng2022recommendation}. 
(2) \textit{Applying LLMs before candidate filtering}. These methods employ LLMs to enrich user/item attributes along with interactions for pre-processing, and then the augmented dataset will be fed into traditional RSs for candidate filtering~\cite{ren2024representation, wei2024llmrec, ren2024enhancing}. However, as shown in Figure~\ref{intro_fig}, both types of methods failed to explore LLMs' capabilities during the candidate filtering process itself, which may lead to suboptimal performance. 

%Moreover, the first category heavily depends on the performance These approaches fail to leverage the powerful reasoning capabilities of LLMs in the retrieval process itself, resulting in suboptimal performance. \textcolor{red}{Moreover, such methods using LLMs for dataset pre-processing, making them unable to handle dynamic user preferences that may evolve over time, which fails to meet the requirements of real-world scenarios.
%Some works treat LLMs as final predictors and directly employ them to generate recommendation results., without directly participating in the candidate filtering process.pre-processing the datasets as data augmenters. Both lines of work failed to explore LLMs' capabilities during the filtering process of candidate items, which may lead to suboptimal performance.

% \begin{figure}
%     \centering
%     \includegraphics[width=\linewidth]{Figs/Intro.pdf}
%     \caption{Comparison among (1) existing methods that use LLMs as predictors for generating the final recommendation results (\textit{e.g.,} TALLRec~\cite{bao2023tallrec}); (2) existing methods that use LLMs as augmenter for data augmentation (\textit{e.g.} LLMRec~\cite{wei2024llmrec}); and (3) our proposed framework that narrows the search space in a stepwise manner, utilizing the reasoning capabilities of LLMs to the retrieval process and seamlessly integrating GNNs for enhanced recommendation performance.}
%     \label{intro_fig}
% \end{figure}%\textcircled{1}

%\textcolor{red}{Candidate filtering starts with a vast pool of all items and selects a critical candidate set that aligns with user information.} 

%p3 
% inspired by工业界通常采用的多轮xx as well as llm推理常用的cot策略，我们提出Chain Of Retrieval ON grAphs，逐步从user-item interaction graph上缩小检索范围进行推荐。
%Inspired by the success of step-by-step reasoning strategies widely adopted in LLM inference~\cite{}, 
In this paper, we propose to leverage LLMs' reasoning abilities during the candidate filtering process, and introduce a coarse-to-fine framework named Chain Of Retrieval ON grAphs (CORONA) for recommendation. As shown in Figure~\ref{subfig3}, \modelname progressively narrows the search space on the user-item interaction graph to improve recommendation accuracy with the help of LLMs. 
% note that llm优势在于其包含的一般性常识/世界知识【需要引用支撑】，因此我们让llm进行粗粒度的偏好和意图推理，而不是从少数具体的item candidates中进行细粒度的选取。
Note that the strength of LLMs lies in their extensive general knowledge and world understanding~\cite{petroni2019language, brown2020language, wu2024survey, zhao2023survey}. %making them particularly effective at answering broad and general questions~\cite{petroni2019language}. 
Therefore, we leverage LLMs for coarse-grained preference and intent reasoning, rather than relying on them for fine-grained selection from similar candidate items. 
% 基于llm推理出的偏好、意图，我们从interaction graph提取出比较相关的subgraph，之后用gnn进行编码并高效地计算相似度xxxx推荐。
%Based on the preferences and intents inferred by the LLM, we extract a relevant subgraph from the interaction graph. This subgraph is then input into a GNN, with the encoded embedding subsequently used to efficiently score all items for recommendation. %We illustrate our framework and its comparison with existing methods .
% Specifically，modelname包含三个阶段的不同粒度的检索：(1)xx(2)xx(3)xx 类似3.2framework overview。

Specifically, \modelname includes three stages of retrieval at different granularities:
(1) We first use the user profile as input to an LLM for preference reasoning, generating a query embedding to perform \textit{preference-assisted retrieval}, which extracts a subgraph aligned with the user's general preferences  from the interaction graph.
(2) We then combine the user's purchase history with statistical information from the previous subgraph, prompting the LLM for intent reasoning. The resulting query embedding supports \textit{intent-assisted retrieval}, refining the subgraph to reflect more personalized and short-term user intent.
(3) Finally, we apply \textit{GNN-enhanced retrieval}, where the GNN processes the subgraph to capture valuable relationships, producing the final recommendation results.
% 再简单说下这么做的好处，比如可以利用到llm的强大推理能力缩小检索范围啥的，又可以利用gnn传统模型进行高效xxx。
In this way, we can leverage the strong reasoning capabilities of LLMs in the item retrieval process, narrowing down relevant items in the entire dataset. Additionally, our framework seamlessly integrates with traditional GNN models to efficiently capture collaborative filtering information, enhancing overall recommendation performance.
% 总结一下实验结果
Extensive experiments on three datasets show that, on average, our model achieves an 18.6\% relative improvement in recall and an 18.4\% relative improvement in NDCG compared to the best baseline, highlighting the effectiveness of this framework.

%p4 列contribution 1提出一种新的利用llm做推荐的思路：让llm先做粗粒度的推理辅助筛选，再用传统模型，可以更好扬长避短充分发挥llm优势 2方法 提出xxx 3实验效果
The contribution of this work are three-fold: 

$\bullet$ We introduce a novel framework for graph-based recommendation, utilizing LLMs to assist coarse-grained retrieval, followed by traditional CF methods for fine-grained recommendation. This allows LLMs to capitalize on their strengths in reasoning and directly involve in the candidate filtering process.% while mitigating the limitations of context length and computational cost. 

$\bullet$ We propose \modelname, a carefully designed three-stage retrieval framework that progressively refines the retrieval process at different levels of granularity. \modelname leverages both the strong capability of LLMs for preference and intent reasoning, as well as typical GNNs for efficient recommendation.

$\bullet$ Extensive experiments on three publicly available datasets demonstrate the effectiveness of our approach for recommendation, with ablation studies validating the necessity of each module.

%CORONA	Chain Of Retrieval ON grAphs
\section{Related Work}
\subsection{Graph-based Recommendation}
Collaborative Filtering is a foundational approach in recommendation systems and has been the subject of extensive research~\cite{chen2020revisiting}. A growing trend in the field is to model user-item interactions as a bipartite graph and apply GNNs to capture high-order collaborative relationships, such as NGCF~\cite{wang2019neural}, LightGCN~\cite{he2020lightgcn}, GraphPro~\cite{yang2024graphpro}, GRCN~\cite{wei2020graph}, PUP~\cite{zheng2020price} and IRGPR~\cite{liu2020personalized}. Specifically, NGCF models the connections between users and items by propagating node embeddings across the graph. LightGCN improves efficiency by eliminating redundant components from the message-passing process in graph-based recommendations. GraphPro incorporates both a temporal prompt mechanism and graph-structural prompt learning into its pre-trained GNN architecture. GRCN dynamically refines the interaction graph structure in response to the model's training progress. PUP designs an encoder with GCN on a predefined heterogeneous graph to capture price awareness, investigating the influence of price feature in ranking stage and enhancing performance. IRGPR proposes a heterogeneous graph to fuse the two information sources, one item relation graph to capture multiple item relationships and one user-item scoring graph to include the initial ranking scores, accomplishing personalized re-ranking with the help of GNN.
In this work, we extend these approaches with LLMs' world knowledge and reasoning ability, thereby improving recommendation performance.

\subsection{LLM-enhanced Recommendation}
Recently, using LLMs to enhance recommendation systems has gained significant attention~\cite{zhao2023recommender, lin2023can, wu2024survey, liu2023pre, liao2024llara, hou2024large, bao2023tallrec, zhang2023recommendation, ma2024xrec, liu2024large, chen2024softmax, kong2024customizing, sun2024large, wang2024reinforcement, shi2024enhancing, zhang2024lorec, yang2024sequential, lin2024data, tan2024towards, geng2024breaking, hua2023tutorial, mysore2023large, sanner2023large, zhang2023chatgpt, kim2024large, zhu2024collaborative, ren2024enhancing, wang2024llmrg}. Existing works can be categorized into the following two types.%, \textit{i.e.,} \textcolor{red}{\textit{applying LLMs after candidate filtering} and \textit{applying LLMs before the candidate filtering}.}

\subsubsection{Applying LLMs after candidate filtering} Some works (\textit{e.g.}~\cite{gao2023chat, hou2024large, wang2023rethinking, kim2024large, zhu2024collaborative}) align recommendation tasks with natural language, and directly prompt LLMs to generate final recommendation results based on a candidate set generated by traditional RSs. Initially, several approaches use pre-trained LLMs and leverage in-context learning capabilities to perform recommendation. For example, Chat-REC~\cite{gao2023chat} constructs a conversational recommender by converting user profiles and interactions into prompts, allowing LLMs to generate recommendations. 
LLMRank~\cite{hou2024large} provides demonstration examples by enhancing the input interaction sequence directly. 
Other methods (\textit{e.g.}~\cite{bao2023tallrec, zhang2023recommendation, zhang2024text, zhang2023collm}) tailor LLMs for recommendation by tuning them with recommendation data. For example, P5~\cite{geng2022recommendation} transforms user interaction data into textual prompts based on item indexes, which are then utilized for language model training. 
InstructRec~\cite{zhang2023recommendation} and TALLRec~\cite{bao2023tallrec} use instructional designs to define recommendation tasks and fine-tune LLMs to follow these instructions for generating recommendations. 
% However, using LLMs directly as predictor faces challenges like limitation of context length and high computational costs. these process information from the entire pool of all items to form a candidate set. This also makes the accuracy of 
{However, these methods heavily depend on the quality of candidate sets generated by traditional approaches~\cite{wei2024llmrec}.}

\subsubsection{Applying LLMs before candidate filtering} Some recent work leverages LLMs for data augmentation as a pre-processing step. Specifically, RLMRec~\cite{ren2024representation} proposes a model-agnostic approach to enhance existing recommenders with LLM generated user profiles. LLMRec~\cite{wei2024llmrec} uses rich online content, including image and text, to augment the interaction graph. In fact, these methods are compatible with our framework for data pre-processing.

We would like to highlight the key difference between these methods and our work: (1) These methods employ LLMs for data augmentation rather than in the retrieval process, making them incapable of handling dynamically changing user intents with LLMs. (2) In contrast, our approach integrates LLM reasoning into the candidate filtering process, allowing instructions to be easily modified to guide recommendations under different contexts. %(2) These methods apply LLMs to augment the entire dataset, which incurs significant overhead when dealing with large and dense graphs, making them unsuitable for real-world recommendation scenarios. Our approach performs LLM reasoning only for the target user, resulting in a more efficient and practical solution tailored to real-world applications.}

%In this work, we take advantage of LLMs' general knowledgeTo address this, our method designs an LLM-assisted retrieval process that fully leverages the reasoning ability and general knowledge of LLMs while enabling scalability and effectiveness.
% llm-based:  however %先讲不调模型直接icl的，再讲会训llm的，然后一起however 逻辑参考intro
%Moreover, LLMs have demonstrated strong commonsense reasoning abilities and excel at general question answering~\cite{petroni2019language}, while their performance may differ on tasks that involve fine-grained ranking within a candidate set. 

%%optional ，此外，llm更擅长利用常识性知识进行一般性粗粒度的推理【需要引用支撑】LLMs do better in general

% In these methods, the reasoning ability of LLMs is not incorporated into the retrieval process of recommendation. Moreover, our approach is compatible with theirs, as they are model-agnostic representation learning methods.
% llm-augmented: compatible%two very recent xxx focus on xxx augmentation, the reasoning ability of llms is not involved into the retrieval process of rec. Besides, compatible

\section{Methodology}
\begin{figure*}
    \centering
    \includegraphics[width=\linewidth]{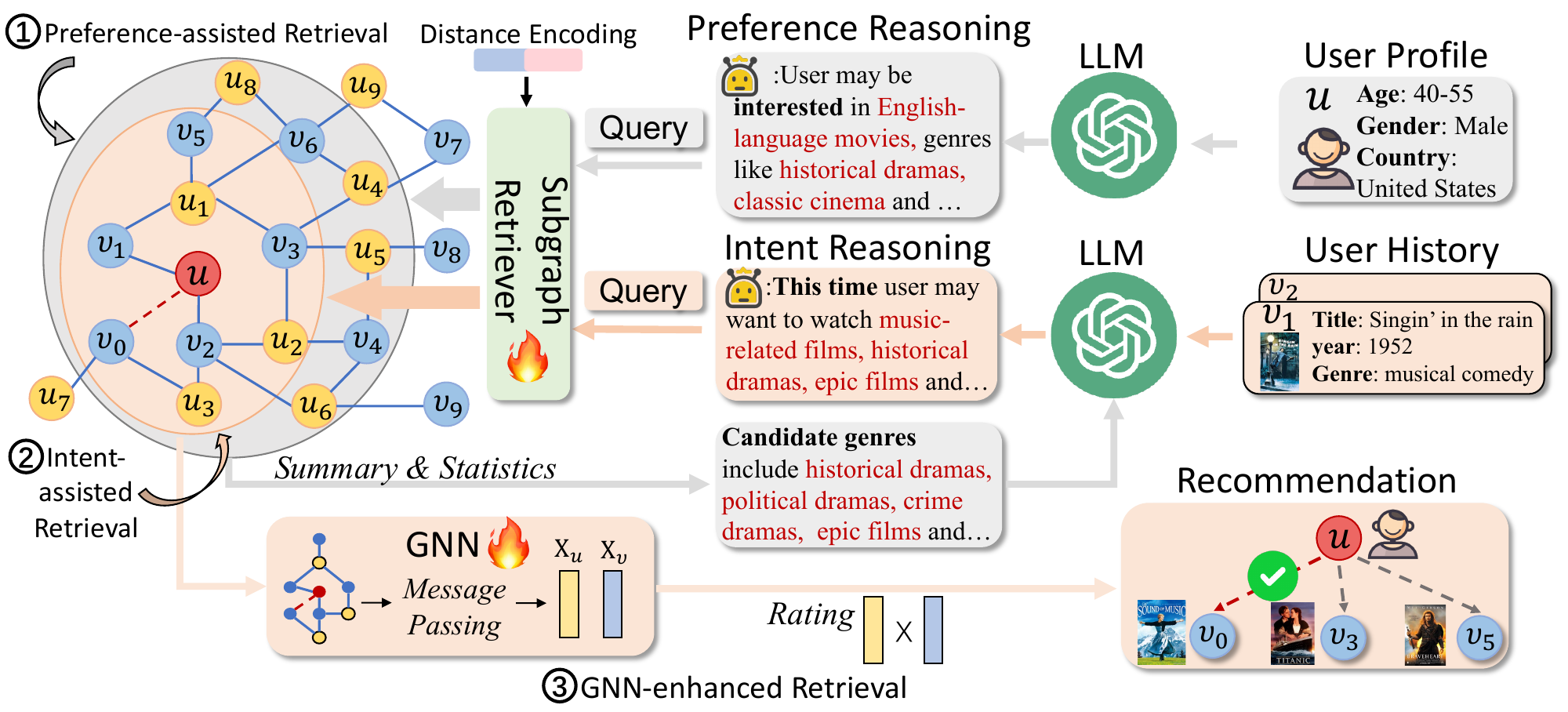}
    \caption{The overall framework of our proposed \modelname, which includes three stages of retrieval from different granularities. Here an LLM is employed to perform coarse-grained preference/intent reasoning based on user profile/history in the first two stages, helping progressively extract a subgraph of relevant users and items from the interaction graph. The final subgraph will be encoded by a GNN for fine-grained recommendation.}
    \label{fig:model}
\end{figure*}
\subsection{Problem Statement}
\subsubsection{Notations}
We focus on graph-based recommendation with textual information \cite{ren2024representation}, where the interaction graph consists of two types of nodes, \textit{i.e.,} users and items. Each node is associated with both textual descriptions and numerical features, and the edges represent purchase records. Formally, we denote the interaction graph as $\mathcal{G} = \left(\mathcal{U}, \mathcal{V}, \boldsymbol{A}\right)$, where $\mathcal{U}$ is the set of users, $\mathcal{V}$ is the set of items and $\boldsymbol{A} \in \{0, 1\}^{(|\mathcal{U}| + |\mathcal{V}|) \times (|\mathcal{U}| + |\mathcal{V}|)}$ represents the adjacency matrix.
% that stores the edges between users and items.
% We follow the method proposed in \cite{} to obtain these textual features. 
%Each user $u$ belongs to the user set $\mathcal{U}$ and each item $v$ belongs to the item set $\mathcal{V}$.
Each user $u \in \mathcal{U}$ has an interaction history $\boldsymbol{L}_u $ as a list of items sorted by interaction time. %Each $\boldsymbol{L}_u$ belongs to the history set $$.\in \boldsymbol{L}_{\mathcal{U}}
Both users and items have corresponding textual information, represented as $\boldsymbol{P}_\mathcal{U}$ and $\boldsymbol{T}_\mathcal{V}$, respectively. 
In addition, feature vectors of users and items can be respectively derived as $\boldsymbol{F}_\mathcal{U}$ and $\boldsymbol{M}_\mathcal{V}$.

\subsubsection{Problem Definition} We follow the settings of recent advances in LLM-augmented recommendation~\cite{wei2024llmrec, ren2024representation} and focus on Top-N Recommendation. Formally, given the interaction graph $\mathcal{G}$ and associate information (including $\boldsymbol{P}_\mathcal{U}$, $\boldsymbol{T}_\mathcal{V}$, $\boldsymbol{F}_{\mathcal{U}}$ and $\boldsymbol{M}_{\mathcal{V}}$), the goal is to predict the next item $v$ that user $u$ will purchase based on interaction history $\boldsymbol{L}_u = \{v_1, v_2, \dots, v_n\}$. Training set $\mathcal{Z}{\text{train}}$ and test set $\mathcal{Z}{\text{test}}$ are both sets of interactions. The adjacency matrix \(\boldsymbol{A}\) ensures that the interactions selected as training and test samples are excluded.

%Consider a classical recommendation task, where each user $u \in \mathcal{U}$ has a textual attribute  $\boldsymbol{P}_u$ and a user feature $\boldsymbol{X}_u$, and each item $v \in \mathcal{V}$ has a textual attribute $\boldsymbol{T}_v$ and an item feature $\boldsymbol{M}_v$. Given a target user $u$ with a interaction history $\boldsymbol{L}_u = \{v_1, v_2, \dots, v_n\}$, the recommender system's goal is to predict the next item $v_{n+1}$. We add the pair $(u, v_{n-1})$ as a training sample to the training set $\mathcal{Z}{\text{train}}$, and add the pair $(u, v_n)$ to the test set $\mathcal{Z}{\text{test}}$. 
% The adjacency matrix $\boldsymbol{A}$ used for training is ensured to exclude the edges $(u, v_{n-1})$ and $(u, v_n)$. 

% The retriever's goal is, given a target user $u$, to identify the most relevant interaction subgraph for this target user denotes as $\mathcal{G}^{\prime} = \left(\mathcal{U}^{\prime}, \mathcal{V}^{\prime}, \boldsymbol{A}^{\prime}\right)$. Here, $\mathcal{U}^{\prime} \subseteq \mathcal{U}, \mathcal{V}^{\prime} \subseteq \mathcal{V}$.
% \boldsymbol{A}^{\prime} = \boldsymbol{A}[U^{\prime}, V^{\prime}]$.

\subsection{Framework Overview}
The framework of the proposed \modelname is shown in Figure~\ref{fig:model}. We conduct three stages of retrieval from different granularities. In the first stage, we take the target user's profile as input, and employ an LLM to infer the user's potential preferences. These preferences are then transformed into query embeddings, and help extract the subgraph from the interaction graph as preference-assisted retrieval. The resulting subgraph is further summarized into natural language. In the second stage, the summary is combined with the user's purchase history for LLM-based intent reasoning. Similarly, we will narrow the scope based on the previous retrieval by extracting a smaller subgraph as intent-assisted retrieval. In the third stage, the final subgraph will be passed to a GNN for message passing, and items are scored by the inner products of user/item embeddings for recommendation.

%Moreover, the  of this step    the fine-grained retrieval.

%of information in a two-step process, retrieving the most relevant subgraph for recommendation.
% It uses user information at different granularities in a two-step retrieval process to gradually narrow down the scope and obtain the most relevant subgraph for recommendation.
%In the coarse-grained retrieval,  and a LLM reasons to generate the user's general preferences.   In the fine-grained retrieval, LLM is further utilized to reason about the user's specific purchase intent,  

\subsection{Preference-assisted Retrieval}
% In our backbone, we use the state-of-the-art GPT-4 for preference reasoning and intent reasoning, generating query content at two different levels of granularity. Next, the query content is further encoded using the OpenAI API. For the retriever, we apply the k-nearest neighbors retrieval method combined with distance encoding. The GNN component is flexible and can be implemented with GCN, GraphTransformer, or other message-passing methods. In the backbone, the retriever and GNN are trained separately with their respective losses. During the inference step, the retriever provides the retrieval results for all users, which are used as the training input for the GNN.
In the first stage, we use an LLM to reason about user preferences based on profiles, helping to extract items and users that align with the target user's interests. 
% Using the results of this reasoning, we can focus on the most relevant parts of the interaction graph, reducing unnecessary noise. 
% By applying an attention-like mechanism, we narrow down the massive interaction graph to a smaller, more focused subgraph, which serves as the foundation for the subsequent fine-grained retrieval. 
The process of preference-assisted retrieval can be broken down into three steps.

\subsubsection{Preference Reasoning}
First, we use an LLM to generate preference reasoning content based on the user's profile, which will further serves as the query embedding.
Specifically, for a target user $u$, the user's profile information $\boldsymbol{P}_u$ includes details such as age, gender, country, and other relevant attributes. We assemble these information into a natural language instruction to input into the LLM. 
The LLM used for preference reasoning can be represented as $\text{LLM}_{PR}(\cdot)$, it is expected to utilize its reasoning ability and stored knowledge to infer the user's preference information and output it in natural language form.
%Since this process relies heavily on the LLM’s reasoning abilities and affects the retrieval quality, we use GPT-4 as the reasoner. 
For instance, in a movie recommendation scenario, if the user's profile $\boldsymbol{P}_u$ indicates she is from France, the LLM should infer that the user might prefer French-language movies in the item set. The instruction for preference reasoning is shown below:
\begin{tcolorbox}[colback=gray!10, colframe=black, rounded corners, boxrule=1.5pt, fontupper=\normalsize, left=2mm, right=2mm, top=1mm, bottom=1mm]
    \textbf{Module: Preference Reasoning} \newline
    \textbf{Instruction:} Please infer the user's movie preferences based on the provided user profile. This may include preferred categories, styles, origins, or years of movies. Begin by summarizing the relevant information, and then provide your preference reasoning predictions using the user profile details along with your general knowledge.\newline
    \textit{User profile:} Age: 25; Gender: Male; Country: US; Language: English; Occupation: Master student...
\end{tcolorbox}
%The CoT (Chain-of-Thought) instruction prompts the LLM to prioritize reasoning over the given information, minimizing the risk of generating hallucinations.

Besides the inferred preference, the LLM's responses may include explanations and reasoning steps, which we believe also reflect the user's preferences in some way. Therefore, we treat all generated content as the preference reasoning result $\boldsymbol{Q}_1 = \text{LLM}_{PR}(\boldsymbol{P}_u)$. Next, $\boldsymbol{Q}_1$ will be encoded to form the query embedding $\boldsymbol{E}_{\boldsymbol{Q}_1} = \text{Encode}(\boldsymbol{Q}_1)$ for subgraph retrieval.

\subsubsection{Subgraph Retriever}
To retrieve the most relevant items based on the query, we design a simple subgraph retriever similar to the attention mechanism. Following the classic collaborative filtering principle~\cite{koren2021advances}, we believe that identifying users with similar interests to the target user will help retrieve more relevant information. While in the interaction graph, users who are closer to each other tend to have more similar interests due to the similarity in their interaction histories.
% For example, we observed that the target user is more likely to notice items purchased by other users who have also bought the same items. 
Therefore, we base the retrieval process on similar users, and focus on those closer to the target user $u$.

Specifically, we first 
% encode the user's textual attributes to generate the users' features $\boldsymbol{X}_{\mathcal{U}}$. These attributes include their historical purchase information $\boldsymbol{A}_{[u]} = \{i_1, i_2, \dots, i_n\}$, with each record detailing the type, style, and year of the purchased items. Next, we
concatenate the distance encoding ${\{\boldsymbol{e}_j\}}_{j=1}^{3}$ to the users' features based on their distances from the target user $u$. 
We use the number of hops between target user $u$ and user $u'$ to represent the distance $\text{dist}(u', u)$ between them.
% We define the distance between users and the target user as the number of hops between them. 
If users \( u \) and \( u' \) have interacted with the same item, they are considered one-hop neighbors, and \( \text{dist}(u', u) = 1 \). Similarly, a user who is a one-hop neighbor of a one-hop neighbor of \( u \) is a two-hop neighbor, with \( \text{dist}(u', u) = 2 \), and so on.
% We use the number of hops between neighbors in the user-only adjacency matrix as the distance, and the user-only adjacency matrix is obtained as follows:
% \begin{equation}
% \label{eq1}
%     \boldsymbol{A}_{\text{user}} = \boldsymbol{A} \cdot \boldsymbol{A}^\top
% \end{equation}
% Here, $\boldsymbol{A}_{\text{user}} \in \{0, 1\}^{|\mathcal{U}| \times |\mathcal{U}|}$, each element \((u_1, u_2)\) in \(\boldsymbol{A}_{\text{user}}\) represents a co-purchases history between users \(u_1\) and \(u_2\).

Our data analysis revealed that very few users in $\mathcal{Z}_{\text{train}}$ pay attention to items interacted by users beyond two-hop neighbors: only 3\% of users on Netflix~\cite{bennett2007netflix} and 0.4\% on MovieLens~\cite{harper2015movielens}. Therefore, users beyond two hops are uniformly assigned $\boldsymbol{e}_3$, aiming to reduce the attention given to these distant users. A linear layer $\text{Linear}_{\theta}$ with trainable parameters $\theta$ is then employed to fuse user embeddings with the distance encoding information and reduce dimensionality.
Specifically, with target user $u$, each $u^{\prime} \in \mathcal{U}$ can get user embedding $\boldsymbol{X}_{u^{\prime}}$ as follows:
\begin{equation}
\label{eq1}
\boldsymbol{X}_{u^{\prime}} = 
    \begin{cases} 
        \text{Linear}_{\theta}({\rm CONCAT}(\boldsymbol{F}_{u^{\prime}}, \boldsymbol{e}_{1})) & \text{if } \text{dist}(u', u) = 1\\
        \text{Linear}_{\theta}({\rm CONCAT}(\boldsymbol{F}_{u^{\prime}}, \boldsymbol{e}_{2})) & \text{if } \text{dist}(u', u) = 2 \\
        \text{Linear}_{\theta}({\rm CONCAT}(\boldsymbol{F}_{u^{\prime}}, \boldsymbol{e}_{3})) & \text{otherwise} 
    \end{cases}
\end{equation}
Finally, we have user embeddings with encoded distance information $\boldsymbol{X}_{\mathcal{U} \setminus \{u\}}$ for all users $\mathcal{U}$ except target user $u$.

%The transformed user embeddings serve as the key in our attention-like retrieval. 
Based on the query generated by the LLM, we use cosine similarity to find the top-$k$ most similar users $\mathcal{U}_1^{\prime}$. Here the target user $u$ is also included. This process is formalized as:
\begin{equation}
\label{eq2}
    % \mathcal{U}_1^{\prime} = \text{argtop}k1_{\{u^{\prime} \in \mathcal{U} \mid u^{\prime} \neq u\}} {\rm cos}(\boldsymbol{E}_{\boldsymbol{Q}_1}, \boldsymbol{X}_{u^{\prime}}) \cup \{u\}
    \mathcal{U}_1^{\prime} = \text{argtop}k_{\{u^{\prime} \in \mathcal{U} \setminus \{u\}\}} {\rm cos}(\boldsymbol{E}_{\boldsymbol{Q}_1}, \boldsymbol{X}_{u^{\prime}}) \cup \{u\}.
\end{equation}
To ensure the extracted subgraph includes the most relevant items, we add all items $\mathcal{V}_1^{\prime}$ connected to $\mathcal{U}_1^{\prime}$ into the subgraph.

\subsubsection{Summary \& Statistics} To assist the intent reasoning in next stage, we summarize the information of the retrieved items $\mathcal{V}_1^{\prime}$ into natural language as auxiliary information. Specifically, we collect the textual information of items in $\mathcal{V}_1^{\prime}$ and then conduct statistics. We include the top 20 most frequent attributes (\textit{e.g.,} movie genres and categories) in the summary as $\text{Summary}(\boldsymbol{T}_{\mathcal{V}_1^{\prime}})$. The example of text template for summarization and statistics is as follows:%, and we take into account the release years of the items for items $v^{\prime} \in \mathcal{V}_1^{\prime}$ for statistical analysis.

%including their categories, genres, and release years. 

%After obtaining the coarse-grained retrieved users $\mathcal{U}_1^{\prime}$, they are used as the candidate set for the fine-grained retrieval. The results from this retrieval are then summarized and fed into the fine-grained reasoner LLM. 
%Since the next reasoning step requires more fine-grained information, such as details about purchased items, 
%The summarized results are then formatted into coherent natural language paragraphs, which serve as part of the input for the fine-grained retrieval. This process can be formally represented as .
% \vspace{-1.3cm}
\begin{tcolorbox}[colback=gray!10, colframe=black, rounded corners, boxrule=1.5pt, fontupper=\normalsize, left=2mm, right=2mm, top=1mm, bottom=1mm]
    \textbf{Module: Summary \& Statistics} \newline
    \textbf{Candidate Genres:} The candidate items belong to the following genres: \textit{[historical drama, political drama, epic film, music-related film, crime drama...]}\newline
    \textbf{Candidate Categories:} The items can be categorized into: \textit{[independent film, documentary, animation ...]}\newline
    \textbf{Years of release:} The movies were released in the following periods: \textit{[1930-1950, 1960-1970, ...]}
\end{tcolorbox}
% \vspace{-1.2cm}

\subsection{Intent-assisted Retrieval}
User intent can be viewed as a more fine-grained and short-term interest compared to preference~\cite{li2021intention}. Therefore, we use the target user's interaction history $\boldsymbol{L}_u = \{v_1, v_2, \dots v_n\}$ as the basis for inferring her intent, and perform the second round of subgraph retrieval. %For example, if the user's interaction history frequently includes historical dramas, it is likely that their next interaction will also lean towards similar content, such as period films or documentaries. leverage the LLM to infer the user's intent.  We , \textit{e.g.,} looking for historical dramas for educational purposes

Specifically, given a target user $u$, we extract the textual information $\boldsymbol{T}_{\boldsymbol{L}_u}$ of the items in $u$'s interaction history, and input them together with $\text{Summary}(\boldsymbol{T}_{\mathcal{V}_1^{\prime}})$ into the LLM for intent reasoning. 

The intent reasoner can be denoted as $\text{LLM}_{IR}(\cdot, \cdot)$, and LLM's response $\boldsymbol{Q}_2 = \text{LLM}_{IR}(\text{Summary}(\boldsymbol{T}_{\mathcal{V}_1^{\prime}}), \boldsymbol{T}_{ \boldsymbol{L}_{u}})$. Similar to the previous step, we encode the response content as query embedding $\boldsymbol{E}_{\boldsymbol{Q}_2} = \text{Encode}(\boldsymbol{Q}_2)$ for intent-assisted retrieval. We use cosine similarity to select the top-${\frac{k}{2}}$ most relevant users as $\mathcal{U}_2^{\prime}$, and include the target user in the user set. 
This process can be formalized as:
\begin{equation}
\label{eq3}
    \mathcal{U}_2^{\prime} = \text{argtop}{\frac{k}{2}}_{\{u^{\prime} \in \mathcal{U}_1^{\prime} \setminus \{u\}\}} {\rm cos}(\boldsymbol{E}_{\boldsymbol{Q}_2}, \boldsymbol{X}_{u^{\prime}}) \cup \{u\}.
\end{equation}
Along with the items $\mathcal{V}_2^{\prime}$ connected to $\mathcal{U}_2^{\prime}$, we have the final retrieved subgraph $\mathcal{G}^{\prime} = (\mathcal{U}_2^{\prime}, \mathcal{V}_2^{\prime}, \boldsymbol{A}[\mathcal{U}_2^{\prime}, \mathcal{V}_2^{\prime}])$.
The instruction example used for intent reasoning is shown below:

\begin{tcolorbox}[colback=gray!10, colframe=black, rounded corners, boxrule=1.5pt, fontupper=\normalsize, left=2mm, right=2mm, top=1mm, bottom=1mm]
    \textbf{Module: Intent Reasoning} \newline
    \textbf{Instruction:} Please infer the user's watching intent based on user history and candidate information. The watching intent represents the user's intention for this viewing, and you should make your selection from within the candidate range. Begin by summarizing the relevant information, then provide your intent reasoning predictions using the user history details, candidate sets, and your general knowledge.\newline
    \textit{User history:} No.1: Title: The Last Emperor; Year: 1987; Genre: History drama; No.2: Singin' in the rain... \newline
    \textit{Candidate summary:} Candidate Genres: ...; Candidate Categories: ...; Years of release: ...
\end{tcolorbox}
% This process can be formally represented as $\boldsymbol{Q}_2 = \text{LLM}(\text{Summary}(\boldsymbol{T}_{\mathcal{I}_1^{\prime}}), \boldsymbol{T}_{i^{\prime} \in \boldsymbol{L}_{u}})$.Here, the CoT instruction encourages the LLM to focus more on reasoning based on the provided information rather than generating hallucinations. 

To learn the parameters in subgraph retriever (\textit{i.e.,} $\theta$ and ${\{\boldsymbol{e}_j\}}_{j=1}^{3}$), we regard the users interacted with ground truth item $v$ as true users, and use the following training loss:
% \begin{equation}
% \label{eq4}
%     \mathcal{L}_1 = - \sum_{u' \in \mathcal{U}} \text{softmax}\left( \text{cos}(\boldsymbol{X}_{u'}, \boldsymbol{M}_v) \cdot \alpha_{u'} \right) \cdot \log\left( \text{cos}(\boldsymbol{X}_{u'}, \boldsymbol{M}_v) \right)
% \end{equation}
\begin{align}
\label{eq4}
    \mathcal{L}_1 = - \sum_{u' \in \mathcal{N}_v} \bigg(& \frac{\exp(\boldsymbol{E}_{\boldsymbol{Q}_1}^{\top} \cdot \boldsymbol{X}_{u'})}{\sum_{u'' \in \mathcal{U}} \exp(\boldsymbol{E}_{\boldsymbol{Q}_1}^{\top} \cdot \boldsymbol{X}_{u''})} \notag \\
    & + \frac{\exp(\boldsymbol{E}_{\boldsymbol{Q}_2}^{\top} \cdot \boldsymbol{X}_{u'})}{\sum_{u'' \in \mathcal{U}} \exp(\boldsymbol{E}_{\boldsymbol{Q}_2}^{\top} \cdot \boldsymbol{X}_{u''})} \bigg),
\end{align}
where $\mathcal{N}_v$ is the set of users connected with item $v$.

% Here, $\text{dist}(u, u^{\prime})$ denotes the number of hops between the target user $u$ and user $u^{\prime}$, as defined previously in the context of distance encoding.

\subsection{GNN-enhanced Retrieval}
Following previous graph-based recommendation methods \cite{he2020lightgcn, wei2024llmrec}, we further introduce a GNN to capture the high-order relations within the retrieved subgraph \(\mathcal{G}^{\prime}\). Formally, a GNN with parameters $\phi$ encodes the target user $u$ based on the extracted subgraph $\mathcal{G}^{\prime}$ as \(\boldsymbol{H}_{u}\). Then we score each item in $G'$ by the inner product between user embeddings and item features:
\begin{equation}
\label{eq6}
    \text{score}(u, v) = \boldsymbol{H}_u^{\top} \cdot \boldsymbol{M}_v.
\end{equation}
Following previous work~\cite{he2020lightgcn}, we use the classical Bayesian Personalized Ranking (BPR) loss for training the GNN. The BPR loss can be calculate by:
\begin{equation}
\label{eq7}
    \mathcal{L}_2 = - \sum_{v' \in \mathcal{V}_\text{neg}} \log \sigma(\text{score}(u, v) - \text{score}(u, v^{\prime})), 
\end{equation}
where $\mathcal{V}_\text{neg}$ is the set of negative items randomly chosen from $\mathcal{V}_2'$. Finally, we select items with the highest scores for top-n recommendation task. 

\section{Experiments}
To validate the effectiveness of our proposed \modelname, we conduct extensive experiments to answer the following research questions (\textbf{RQ}s): \textbf{RQ1:} How effective is our proposed \modelname compared to the state-of-the-art baselines? \textbf{RQ2:} Is the proposed \modelname also effective for recommending items in cold-start setting? \textbf{RQ3:}Has each component of our framework played its role effectively? \textbf{RQ4:} How do different values of key parameters influence the method's performance? \textbf{RQ5:} How efficient is CORONA compared with previous methods? \textbf{RQ6:} Does the LLM-based reasoning process in \modelname offer some interpretability? 

\subsection{Experimental Setup}
\subsubsection{Datasets}
We perform experiments on three publicly available datasets, \textit{i.e.,} Netflix\footnote{https://www.kaggle.com/datasets/netflix-inc/netflix-prize-data}, MovieLens\footnote{https://files.grouplens.org/datasets/movielens/ml-10m-README.html} and Amazon-book\footnote{https://cseweb.ucsd.edu/~jmcauley/datasets/amazon/links.html}.
% To ensure a fair comparison, we provide only textual information as side information for all methods. 
For baselines that cannot directly utilize textual information, such as LightGCN~\cite{he2020lightgcn}, we use text encodings as node features.
% We follow existing work to split the train, validation, and test sets.
For the Netflix and MovieLens datasets, we use the same split as in LLMRec~\cite{wei2024llmrec}; for the Amazon-book dataset, we follow the split from RLMRec~\cite{ren2024representation}.

$\bullet$ Netflix dataset~\cite{bennett2007netflix} is sourced from the Kaggle website. 
% We derive the interaction graph $\mathcal{G}$ and user interaction history $\boldsymbol{L}_{\mathcal{U}}$ from the users' viewing history. For items, we combine the movie title, year, genre, and categories as textual information, denoted as $\boldsymbol{T}_{\mathcal{V}}$, while for users, we use age, gender, favorite directors, country, and language as textual information, denoted as $\boldsymbol{P}_{\mathcal{U}}$. Additionally, 
We use BERT~\cite{reimers2019sentence} to encode the textual information of users and items, obtaining user features $\boldsymbol{F}_{\mathcal{U}}$ and item features $\boldsymbol{M}_{\mathcal{V}}$, respectively.

$\bullet$ MovieLens dataset~\cite{harper2015movielens} is sourced from ML-10M.
% and we obtain the interaction graph $\mathcal{G}$ and user interaction history $\boldsymbol{L}_{\mathcal{U}}$ from the users' viewing history. The side information, including movie title, year, and genre, is combined as the item textual feature $\boldsymbol{T}_{\mathcal{V}}$, while user age, gender, country, and language are combined as the user textual feature $\boldsymbol{P}_{\mathcal{U}}$. Additionally, 
We encode these textual information using BERT~\cite{reimers2019sentence} as features $\boldsymbol{F}_{\mathcal{U}}$ and $\boldsymbol{M}_{\mathcal{V}}$.

$\bullet$ Amazon-book dataset~\cite{ni2019justifying} contains book review records from 2000 to 2014. 
% We treat each review as a user-item interaction and derive the interaction graph $\mathcal{G}$ and user interaction history $\boldsymbol{L}_{\mathcal{U}}$ from the records. We use the book title, year, and categories as the textual information for items $\boldsymbol{T}_{\mathcal{V}}$, and the user's review content as the textual information for users $\boldsymbol{P}_{\mathcal{U}}$. These 
Information are encoded by BERT~\cite{reimers2019sentence} to obtain features $\boldsymbol{F}_{\mathcal{U}}$ and $\boldsymbol{M}_{\mathcal{V}}$.

%For baselines that typically require multimodal data, such as LLMRec, we excluded inputs like images and other non-textual modalities. For baselines that cannot directly utilize textual information, such as LightGCN, we used text encodings as node features.
\subsubsection{Evaluation Protocols}
Following previous work~\cite{wang2020disenhan, he2020lightgcn, wei2024llmrec}, we evaluate our approach in the top-K item recommendation task using two common metrics: Recall (R@$K$) and Normalized Discounted Cumulative Gain (N@$K$), where $K$ is set to $10$, $20$, and $50$. We employ the all-ranking strategy, and report averaged results from five independent runs.%, setting K to  (reasonable for all-ranking). Statistical significance analysis is conducted by calculating P-values against the best-performing baseline. 
\subsubsection{Methods for Comparison}
To fully demonstrate the effectiveness of our proposed \modelname, we compare a number of baselines from three groups. 
(1) Graph-based Collaborative Filtering Methods: These approaches leverage GNNs to capture the structural relationships in the interaction graph, including NGCF~\cite{wang2019neural}, LightGCN~\cite{he2020lightgcn}, GraphPro~\cite{yang2024graphpro} and GRCN~\cite{wei2020graph}.
% NGCF~\cite{} models the connections between users and items by propagating node embeddings across the graph. LightGCN~\cite{} improves efficiency by eliminating redundant components from the message-passing process in graph-based recommendations. GraphPro~\cite{} incorporates both a temporal prompt mechanism and graph-structural prompt learning into its pre-trained GNN architecture. GRCN~\cite{} dynamically refines the interaction graph structure in response to the model's training progress. 
(2) LLM for Recommendation: These methods apply LLMs to recommendation tasks to improve performance metrics, including TALLRec~\cite{bao2023tallrec}, BinLLM~\cite{zhang2024text}, RLMRec~\cite{ren2024representation}, LLMRec~\cite{wei2024llmrec}. We also compare CORONA with an alternative LLM retrieval method: G-retriever~\cite{he2024g}, following its original approach described in the paper to perform subgraph construction and then recommend items based on similarity scoring.

For the experiments under item cold-start setting, we select several methods specifically designed for cold-start scenarios along with the ``LLM for Recommendation'' methods as baselines. The selected baselines include DropoutNet~\cite{volkovs2017dropoutnet}, ALDI~\cite{huang2023aligning}, TALLRec~\cite{bao2023tallrec}, BinLLM~\cite{zhang2024text}, RLMRec~\cite{ren2024representation}, LLMRec~\cite{wei2024llmrec}, LLM-Ins~\cite{huang2024large}.
% Cold-start methods: (1) DropoutNet~\cite{} trains a network designed for cold-start by applying dropout to simulate missing data.  
% (2) ALDI~\cite{} transfers behavioral information from warm items to cold items by leveraging their interactions. (3) LLM-Ins~\cite{} simulates the vivid interactions and teansform them directly. ``LLM for Rec'' methods: TALLRec, BinLLM, RLMRec and LLMRec.

\subsubsection{Implementation Details}
We follow existing methods \cite{wei2024llmrec} to obtain the textual attribute of users $\boldsymbol{P}_{\mathcal{U}}$ and items $\boldsymbol{T}_{\mathcal{V}}$, and derive 128-dimensional features for both users $\boldsymbol{F}_{\mathcal{U}}$ and items $\boldsymbol{M}_{\mathcal{V}}$ as well. We employ OpenAI's ``GPT-4o-mini'' for preference and intent reasoning and the responses are encoded using OpenAI's ``text-embedding-ada-002,'' also with 128-dimensional outputs. The distance encoding \( \{\boldsymbol{e}\}_{j=1}^3 \) is set to 2 dimensions. The linear layer $\text{Linear}_{\theta}$ use a 130 $\times$ 128-dimensional linear layer. For the main experiment with the GCN method, we use a two-layer GCN  with hidden dimension set at 128. GraphTransformer is used as provided in the original paper~\cite{yun2019graph}. We set the size of the negative set to 10 for Eq~\ref{eq7}, and employ the Adam optimizer with a learning rate of 1e-6 for parameter training. Early stopping with a patience setting of 10 steps is also used during training. All experiments are conducted on an A800 GPU with 80GB of memory.
%The baseline method ``GCN (fixed 1-hop)'' uses one message-passing layer and two linear layers to maintain parameter parity.

\subsection{Main Results (RQ1)}
To answer \textbf{RQ1}, we conduct recommendation experiments with results shown in Table~\ref{main_table}. From the results, we can see that: (1) Our method \modelname with LLM-empowered reasoning consistently outperforms the state-of-the-art baselines on all three datasets. On average, \modelname has 17.66\% relative improvement in recall and 16.06\% in NDCG compared to the best baseline. This improvement showcases the effectiveness of our framework. 
% (2) Among the GCN-based and GT-based methods, our LLM-based subgraph retrieval demonstrates a clear advantage over others. While the same GNN is applied, our method relatively yields a higher recall by 110.73\% and a higher NDCG by 112.03\% on average. This highlights that our designed retrieval process significantly contributes to more accurate recommendations. 
(2) The LLM-based recommendation methods are the most competitive baselines, as they also leverage LLMs' capabilities to enhance recommendation. But our approach still show a relative improvement of 30.67\% in recall and 31.24\% in NDCG on average. %(4) Notably, to ensure a fair comparison, we provided only textual information as side information for all methods. For baselines that typically require multimodal data, such as LLMRec, we excluded inputs like images and other non-textual modalities. For baselines that cannot directly utilize textual information, such as LightGCN, we used text encodings as node features.% (2) In terms of significance, our experimental results show a statistically significant improvement in most cases, where the p-value is less than 0.05, except for two metrics: R@50 on the Netflix dataset and N@20 on the MovieLens dataset. 

\renewcommand{\arraystretch}{1.5}
\setlength{\tabcolsep}{3pt}
\begin{footnotesize}
\begin{table*}[ht]
\caption{Recommendation performance on three datasets in terms of \textit{Recall@}10/20/50, and \textit{NDCG@}10/20/50.}
\vspace{-0.4cm}
\centering
    \begin{tabular}{@{} c|>{\columncolor{gray!30}}cc>{\columncolor{gray!30}}cc>{\columncolor{gray!30}}cc|>{\columncolor{gray!30}}cc>{\columncolor{gray!30}}cc>{\columncolor{gray!30}}cc|>{\columncolor{gray!30}}cc>{\columncolor{gray!30}}cc>{\columncolor{gray!30}}cc @{} }
    \Xhline{0.8pt}
        Datasets & \multicolumn{6}{c|}{Netflix} &  \multicolumn{6}{c|}{MovieLens} & \multicolumn{6}{c}{Amazon-book}	  \\ 
        \hline
        Methods  & R@10 & N@10 & R@20 & N@20 & R@50 & N@50 & R@10 & N@10 & R@20 & N@20 & R@50 & N@50 & R@10 & N@10 & R@20 & N@20 & R@50 & N@50 \\ \hline
        NGCF & 0.0357 & 0.0163 & 0.0693 & 0.0231 & 0.1090 & 0.0335 & 
        0.1153	&0.1007&	0.1305&	0.1164&	0.1636	&0.1261&
        0.0844	&0.0637	&0.1294	&0.0783&	0.2151&	0.1023
        \\ 
         LightGCN & 0.0385 & 0.0152 & 0.0661 & 0.0222 & 0.1252 & 0.0336 &0.0966	&0.1013&	0.1314	&0.1173	&0.1647	&0.1297&
         0.0857	&0.0646&	0.1303&	0.0791&	0.2154&	0.1028 \\ 
         GraphPro & 0.0397 & 0.0174 & 0.0701 & 0.0235 & 0.1263 & 0.0350 
         &0.1376&	0.1121&	0.1462&	0.1296&	0.1757&	0.1411
         &0.0865	&0.0656	&0.1345	&0.0839	&0.2203&	0.1085\\ 
         GRCN & 0.0392 & 0.0169 & 0.0695 & '0.0260 & 0.1318 & 0.0348 & 0.1175&	0.1017&	0.1356&	0.1192&	0.1654&	0.1308 &
         0.0862&	0.0653&	0.1332	&0.0826	&0.2179	&0.1067\\ \hline

         TALLRec	 & 0.0498&	0.0251&	0.0795	&0.0338	&0.1303&	0.0420 &0.2521&	0.1104&	0.3522&	0.1748&	0.5017&	0.1921 
        & 0.0857	&0.0729&	0.1277&	0.0838&	0.2420&	0.1169\\ 
         BinLLM	&0.0502&	0.0254&	0.0797	&0.0340	&0.1312&0.0431 & 0.2584&	0.1187&	0.3657&	0.1771&	0.5094&	0.1995&
         \underline{0.0972}&	\underline{0.0741}&	\underline{0.1512}&	\underline{0.0939}&	\underline{0.2451}&	\underline{0.1194}\\ 
         
         RLMRec	&0.0504&	0.0257&	0.0806&	0.0341&	0.1314&	0.0439
         &\underline{0.2595}&	\underline{0.1195}&	\underline{0.3668}&	\underline{0.1779}&	\underline{0.5108}&	\underline{0.2001}
         &0.0968&	0.0738&	0.1499&	0.0913&	0.2432&	0.1175 \\
         LLMRec	&\underline{0.0526}	&\underline{0.0271}&	\underline{0.0808}&	\underline{0.0342}&	\underline{0.1317}&\underline{ 0.0442}
         &0.2582	&0.1187&	0.3528	&0.1750	&0.5050&	0.1904
         &0.0862	&0.0733	&0.1285	&0.0840	&0.2424	&0.1171 \\ 
         % \hline

%          GCN (full graph)&

%          0.0307&	0.0179&	0.0662&	0.0217&	0.1012&	0.0247&
%         0.1056&	0.0808&	0.1371&	0.0993&	0.1600&	0.1208&
%         0.0786&	0.0405&	0.1121&	0.0556&	0.1845&	0.0797\\
    
% GCN (fixed 1-hop)&
% 0.0319&	0.0107&	0.0617&	0.0169&	0.0981&	0.0217&
% 0.1123&	0.0804&	‘0.1320&	0.1037&	0.1599&	0.1181&
% 0.0773&	0.0427&	0.1133&	0.0516&	0.1925&	0.0790\\

% GCN (fixed 2-hop)&
% 0.0305&	0.0097&	0.0632&	0.0184&	0.0987&	0.0274&
% 0.1092&	0.0912&	0.1244&	0.1086&	0.1495&	0.1178&
% 0.0726&	0.0443&	0.1135&	0.0582&	0.1882&	0.0796\\

G-Retriever&
0.0327&	0.0134&	0.0678&	0.0213&	0.1106&	0.0323&
0.1179&	0.1052&	0.1294&	0.1121&	0.1569&	0.1225&
0.0831&	0.0616&	0.1248&	0.0719&	0.2123&	0.0969\\
\hline

        \modelname & 
        \textbf{0.0616}&	\textbf{0.0279}&	\textbf{0.0938}&	\textbf{0.0416}&	\textbf{0.1452}& \textbf{0.0487}&
        \textbf{0.3033}&	\textbf{0.1565}&	\textbf{0.4214}&	\textbf{0.2017}&	\textbf{0.5745}&	\textbf{0.2507}&
        \textbf{0.1206}&	\textbf{0.0855}&	\textbf{0.1857}&	\textbf{0.1089}&	\textbf{0.3048}&	\textbf{0.1299}
        \\
         \hline
         
        % Improv. & 19.39\% &17.34\% &	16.09\%	&21.64\%&10.25\%	&10.18\% 
        %  &18.38\%&30.96\%&14.89\%&13.38\%&12.47\%&30.88\%
        %  &24.07\%&15.38\%&27.51\%&17.36\%&24.36\%&8.79\%
         Improv. & 17.11\% &2.95\% &	16.09\%	&21.64\%&10.25\%	&10.18\% 
         &16.88\%&30.96\%&14.89\%&13.38\%&12.47\%&25.29\%
         &24.07\%&15.38\%&22.82\%&15.97\%&24.36\%&8.79\%
         \\ 
         \Xhline{0.8pt}
    \end{tabular}
\label{main_table}
\end{table*}
\end{footnotesize}

\subsection{Cold-start Results (RQ2)}
% Item cold-start is a long-standing challenge for recommendation systems.
To answer \textbf{RQ2}, we focus on items with no more than two interactions, and set up an item cold-start scenario for evaluation. From the results shown in Table~\ref{cold_start}, we can observe that: (1) Our method outperforms all baselines, with an average relative improvement of 8.24\% in recall and 10.68\% in NDCG. This demonstrates that our framework remains high utility in cold-start scenarios by leveraging the reasoning capabilities of LLMs to mitigate data sparsity. (2) Note that LLM-Ins~\cite{huang2024large} is specialized for only cold-start settings, and performs the best among baselines. While our approach can outperform LLM-Ins across all datasets and metrics, showcasing the value of our framework. (3) The performance of traditional cold-start methods, namely DropoutNet and ALDI, is relatively low, highlighting the advantage of using LLMs to process textual data. Our method relatively improves recall by 35.51\% over traditional methods and 10.39\% over other LLM baselines, and NDCG by 37.48\% over traditional methods and 11.62\% over other LLM baselines.

\renewcommand{\arraystretch}{1.5}
\setlength{\tabcolsep}{3pt}
\begin{footnotesize}
\begin{table*}[ht]
\caption{Recommendation performance under item cold-start setting in terms of \textit{Recall@}10/20/50, and \textit{NDCG@}10/20/50.}
\vspace{-0.4cm}
\centering
    \begin{tabular}{c|>{\columncolor{gray!30}}cc>{\columncolor{gray!30}}cc>{\columncolor{gray!30}}cc|>{\columncolor{gray!30}}cc>{\columncolor{gray!30}}cc>{\columncolor{gray!30}}cc|>{\columncolor{gray!30}}cc>{\columncolor{gray!30}}cc>{\columncolor{gray!30}}cc}
    \Xhline{0.8pt}
        Datasets & \multicolumn{6}{c|}{Netflix} &  \multicolumn{6}{c|}{MovieLens} & \multicolumn{6}{c}{Amazon-book}	  \\ 
        \hline
        Methods & R@10 & N@10 & R@20 & N@20 & R@50 & N@50 & R@10 & N@10 & R@20 & N@20 & R@50 & N@50 & R@10 & N@10 & R@20 & N@20 & R@50 & N@50 \\ \hline
        DropoutNet & 0.0175&	0.0083&	0.0323&	0.0157&	0.0414&	0.0195 &
         0.1352	&0.0538	&0.2107	&0.1101	&0.3622	& 0.1495&
         0.0425	&0.0264	&0.0832	&0.0496	&0.1482	&0.1007
          \\ 
         ALDI & 
        0.0247	&0.0099&	0.0409&	0.0161	&0.0527	&0.0202 & 
      0.1436	&0.0552	&0.2244	&0.1102&	0.3613	&0.1488&
      0.0525	&0.0307	&0.0898	&0.0517	&0.1511	&0.1026
         \\ 
         TALLRec & 
         0.0124&	0.0075&	0.0301&	0.0152	&0.0322	&0.0181&
         0.1224	&0.0525	&0.2083	&0.1099	&0.3597&	0.1475&
         0.0459	&0.0278	&0.0732	&0.0472&	0.1368	&0.1004\\
         
         BinLLM & 
         0.0218	&0.0086	&0.0395	&0.0158	&0.0513	&0.0200&
         0.01373	&0.0547	&0.2159	&0.1095	&0.3638&	0.1477&
         0.0514	&0.0296	&0.0873	&0.0508	&0.1485	&0.1012\\
      RLMRec & 
      0.0243	&0.0096	&0.0412	&0.0164	&0.0535	&0.0207&
      0.1576	&0.0588	&0.2473	&0.1105	&0.3626&	0.1501&
      0.0571	&0.0319	&0.0907	&0.0526	&0.1524	&0.1035\\
         LLMRec & 
         0.0119&	0.0073	&0.0296&	0.0142&	0.0315	&0.0175&
         0.1163	&0.0517	&0.2052	&0.1095	&0.3594	&0.1473&
         0.0454	&0.0273	&0.0725	&0.0464&	0.1357	&0.0971\\
         LLM-Ins &
         \underline{0.0281}	&\underline{0.0126}	&\underline{0.0490}	&\underline{0.0182}&	\underline{0.0739}	&\underline{0.0195}&
        \underline{ 0.1826}	&\underline{0.0854}	&\underline{0.2689}	&\underline{0.1372}&	\underline{0.4063}	&\underline{0.1665}&
        \underline{0.0875}	&\underline{0.0417}	&\underline{0.1167}	&\underline{0.0704}&	\underline{0.1645}&	\underline{0.1206}
         \\ \hline
       
    \rule{0pt}{12pt} \makecell{\modelname}& 
    \textbf{0.0295}	& \textbf{0.0130}	&\textbf{0.0519}	& \textbf{0.0201}	& \textbf{0.0791}&	 \textbf{0.0217}&
    \textbf{0.2054}	&\textbf{0.0933}	&\textbf{0.2917}	&\textbf{0.1425}	&\textbf{0.4582}&	\textbf{0.1931}&
    \textbf{0.0916}	&\textbf{0.0453	}&\textbf{0.1212}	&\textbf{0.0764}	&\textbf{0.1788}&	\textbf{0.1314}\\
    % \hline
    % \rule{0pt}{12pt} \makecell{\modelname \\ with GT}&
    % 0.0288	&0.0128&	\textbf{0.0524}&	0.0194	&0.0777	&0.0209&
    % 0.1971	&\textbf{0.0949	}&0.2911	&\textbf{0.1502	}&0.455&	0.1893&
    % \textbf{0.0952}	&0.0447	&0.1209	&0.0757	&0.1751&	0.1274
         
         % \\ 
    \hline
    % Improv.&4.98\% & 3.17\%& 6.94\%& 10.44\%& 7.04\%& 11.28\%& 12.49\%& 11.12\%& 8.48\%& 9.48\%& 12.77\% & 15.98\%& 8.80\%& 8.63\%& 3.86\%&8.52\%&8.69\%&8.96\% \\
    Improv.&4.98\% & 3.17\%& 5.91\%& 10.44\%& 7.04\%& 11.28\%& 12.49\%& 9.25\%& 8.48\%& 3.86\%& 12.77\% & 15.98\%& 4.69\%& 8.63\%& 3.86\%&8.52\%&8.69\%&8.96\% \\
         \Xhline{0.8pt}
    \end{tabular}
\vspace{-0.2cm}
\label{cold_start}
\end{table*}
\end{footnotesize}

\subsection{Ablation Study (RQ3)}
To answer \textbf{RQ3}, we consider three categories of ablated variants to test whether each design of CORONA is effective and necessary. 
% We present the results in Table~\ref{different_combination} and Table~\ref{ablation_study}.

\subsubsection{Combinations of Different Subgraph Retrieval Methods and GNNs} To show the effectiveness of our proposed LLM-based subgraph retrieval, we consider three other methods for extracting relevant subgraphs: full graph, 1-hop neighbors of target user (fixed 1-hop) and 2-hop neighbors of target user (fixed 2-hop). Also, to show that our \modelname is compatible with different GNNs, we test 2-layer GCN~\cite{kipf2016semi} and GraphTransformer (GT)~\cite{yun2019graph} for recommendation based on extracted subgraphs. We present the results in Table~\ref{different_combination}. From the results, we observe that: among the GCN-based and GT-based methods, our LLM-based subgraph retrieval demonstrates a clear advantage over others. While the same GNN is applied, our method relatively yields a higher recall by 110.73\% and a higher NDCG by 112.03\% on average. This highlights that our designed retrieval process significantly contributes to more accurate recommendations. Besides, GT-based CORONA yields competitive performance with GCN-based one, showing the compatibility of our proposed framework.

\subsubsection{Ablation of Different Components} 
% For ``w/o LLM-assisted Retrieval'', we remove the whole subgraph retrieval part and use GNN-enhanced retrieval for recommendation. 
We ablate each component of CORONA individually to test the effectiveness of the design. The results are shown in Table~\ref{ablation_study}. For ``w/o Preference-assisted Retrieval'' and ``w/o Intent-assisted Retrieval'', we remove the preference-assisted and intent-assisted retrieval components respectively, and use the remaining LLM-assisted retrieval combined with GNN-enhanced retrieval for recommendation. For ``w/o GNN-enhanced Retrieval'', we remove the GNN module $\text{GNN}_\phi$ that updates the embedding of target user by message passing. For ``w/o Preference Reasoning'' , we remove the preference reasoner $\text{LLM}_{PR}$ and directly encode user profile to form the query embedding $\boldsymbol{E}_{\boldsymbol{Q}_1}$. Similarly, for ``w/o Intent Reasoning'', we encode textual user history to form query embedding $\boldsymbol{E}_{\boldsymbol{Q}_2}$. 

From the results, we observe that: (1) The full model \modelname always yields the best performance, demonstrating the soundness of our design and the necessity of each component. 
Compared to the average of variants, the full framework achieves a relative improvement of 20.21\% in recall and 19.52\% in NDCG.
(2) The "w/o Preference Reasoning" and "w/o Intent Reasoning" variants have the weakest performance, highlighting the importance of LLM reasoning in our framework. Note that the above two variants have even worse performance than the ``w/o Preference-assisted Retrieval'' and ``w/o Intent-assisted Retrieval'' variants, indicating that LLM's reasoning abilities and general knowledge indeed contribute effectively to the recommendation task. (3) The variants without intent reasoning/assistance perform worse than those without preference reasoning/assistance. This observation highlights the importance of using LLMs to handle dynamic user intents, whereas previous methods using LLMs for pre-processing are unable to infer such dynamic information effectively.
% preference v.s. intent, llm dynamic necessary, pre-processing unable

\subsubsection{Influence of Different LLMs}
\begin{figure}
    \centering
    \includegraphics[width=\linewidth]{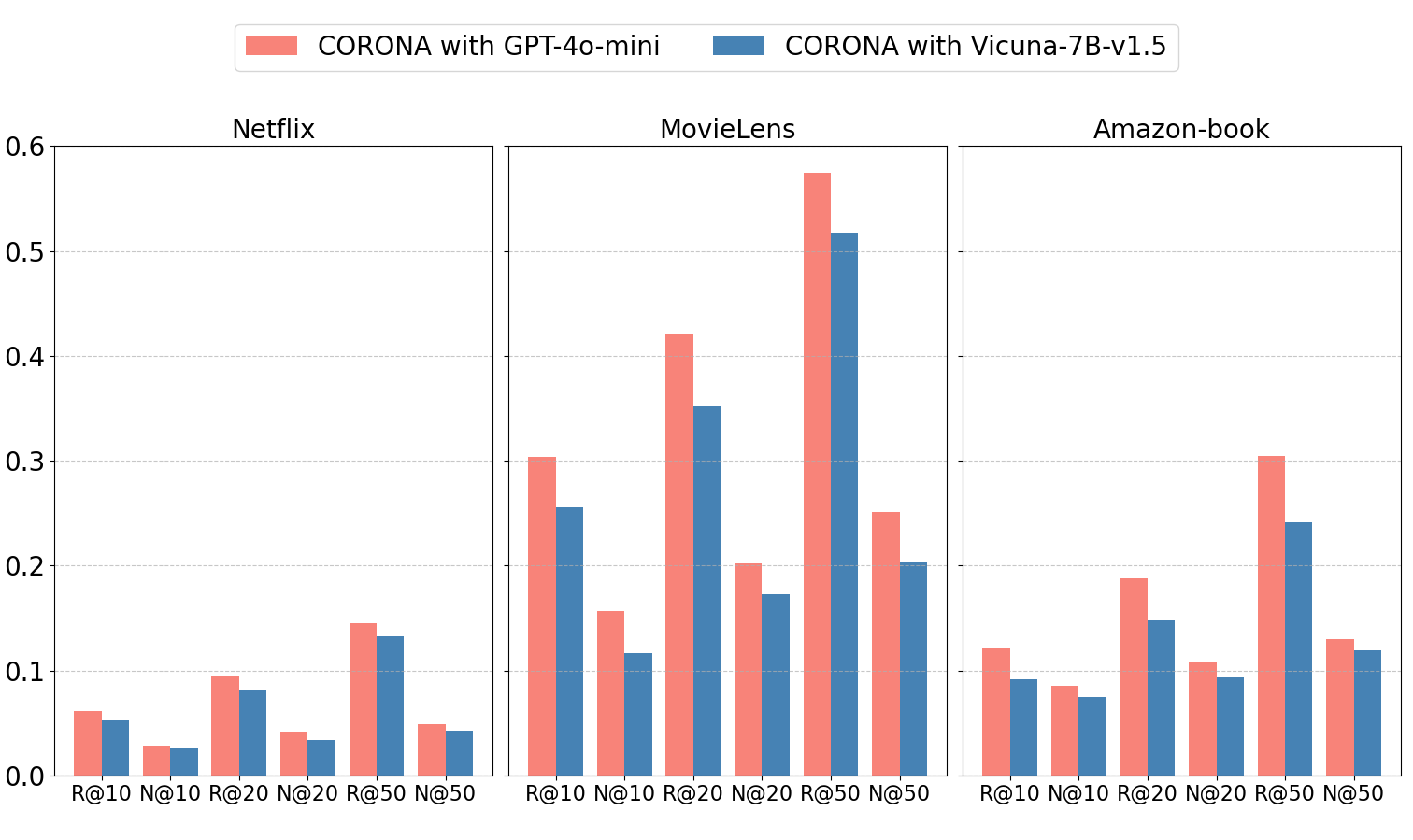}
    \vspace{-0.8cm}
    \caption{Applying different LLMs in \modelname on three datasets in terms of \textit{Recall@}10/20/50 and \textit{NDCG@}10/20/50.}
    \label{llm_comparison}
\vspace{-0.5cm}
\end{figure}

We also investigate the impact of using different LLMs on model performance, with results shown in Figure~\ref{llm_comparison}. Besides GPT-4o-mini used in main experiments, we also test CORONA with a small open-source LLM, \textit{i.e.,} Vicuna-7B-v1.5~\cite{chiang2023vicuna}. Vicuna is an open-source chatbot developed by fine-tuning LLaMA~\cite{touvron2023llama} with conversations shared by users. We deploy the Vicuna-7B-v1.5 model locally for preference and intent reasoning, using the same "text-embedding-ada-002" encoder to ensure the only difference lies in the reasoning content. The results show that CORONA with Vicuna-7B-v1.5 is already slightly better than previous methods, while CORONA with GPT-4o-mini further yields an average improvement of 15.87\%. This demonstrates that higher-quality reasoning leads to more significant performance gains.

\renewcommand{\arraystretch}{1.5}
\setlength{\tabcolsep}{2.5pt}
\begin{footnotesize}
\begin{table*}[ht]
\caption{Combinations of different subgraph retrieval methods and GNNs on three datasets in terms of \textit{Recall@}10/20/50, and \textit{NDCG@}10/20/50.}
\vspace{-0.4cm}
\centering
    \begin{tabular}{@{} c|>{\columncolor{gray!30}}cc>{\columncolor{gray!30}}cc>{\columncolor{gray!30}}cc|>{\columncolor{gray!30}}cc>{\columncolor{gray!30}}cc>{\columncolor{gray!30}}cc|>{\columncolor{gray!30}}cc>{\columncolor{gray!30}}cc>{\columncolor{gray!30}}cc @{} }
    \Xhline{0.8pt}
        Datasets & \multicolumn{6}{c|}{Netflix} &  \multicolumn{6}{c|}{MovieLens} & \multicolumn{6}{c}{Amazon-book}	  \\ 
        \hline
        Variants  & R@10 & N@10 & R@20 & N@20 & R@50 & N@50 & R@10 & N@10 & R@20 & N@20 & R@50 & N@50 & R@10 & N@10 & R@20 & N@20 & R@50 & N@50 \\ \hline
GCN (full graph)&
0.0307&	\underline{0.0179}&	\underline{0.0662}&	\underline{0.0217}&	\underline{0.1012}&	0.0247&
0.1056&	0.0808&	\underline{0.1371}&	0.0993&	\underline{0.1600}&	\underline{0.1208}&
\underline{0.0786}&	0.0405&	0.1121&	0.0556&	0.1845&	\underline{0.0797}\\
    
GCN (fixed 1-hop)&
\underline{0.0319}&	0.0107&	0.0617&	0.0169&	0.0981&	0.0217&
\underline{0.1123}&	0.0804&	‘0.1320&	0.1037&	0.1599&	0.1181&
0.0773&	0.0427&	0.1133&	0.0516&	\underline{0.1925}&	0.0790\\

GCN (fixed 2-hop)&
0.0305&	0.0097&	0.0632&	0.0184&	0.0987&	\underline{0.0274}&
0.1092&	\underline{0.0912}&	0.1244&	\underline{0.1086}&	0.1495&	0.1178&
0.0726&	\underline{0.0443}&	\underline{0.1135}&	\underline{0.0582}&	0.1882&	0.0796\\

GCN (CORONA) & 
\textbf{0.0616}&\textbf{0.0279}&	\textbf{0.0938}&	\textbf{0.0416}&	\textbf{0.1452}& \textbf{0.0487}&
\textbf{0.3033}&	\textbf{0.1565}&	\textbf{0.4214}&	\textbf{0.2017}&	\textbf{0.5745}&	\textbf{0.2507}&
\textbf{0.1206}&	\textbf{0.0855}& \textbf{0.1857}& \textbf{0.1089}&	\textbf{0.3048}&	\textbf{0.1299}
\\
\hline

GT (full graph)&
0.0302 &0.0104&	0.0641&	\underline{0.0192}&	0.1008&	0.0285&
\underline{0.1055}&	0.0907&	0.1235& \underline{0.1074}&	0.1482&	0.1170&
\underline{0.0719}&	0.0435&	0.1127&	0.0571&	0.1854&	0.0787\\
GT (fixed 1-hop)&
0.0294&	0.0081&	0.0655&	0.0152&	0.0988&	\underline{0.0342}&
0.1035&	\underline{0.1039}&	0.1170&	0.0828&	\underline{0.1501}&	0.1137&
0.0672&	0.0330&	0.1109&	0.0451&	0.1835&	0.0787\\
GT (fixed 2-hop)&
\underline{0.0345}&	\underline{0.0117}&	\underline{0.0665}&	0.0160&	\underline{0.1119}&	0.0305&
0.1044&	0.0864&	\underline{0.1370}&	0.1043&	0.1463&	\underline{0.1193}&
0.0696&	\underline{0.0453}&	\underline{0.1187}&	\underline{0.0601}&	\underline{0.1855}&	\underline{0.0790}\\

% G-Retriever with GT&
% 0.0335&	0.0141&	0.0652&	0.0204&	0.1127&	0.0344&
% 0.1102&	0.1091&	0.1281&	0.1107&	0.1557&	0.1213&
% 0.0873&	0.0638&	0.1261&	0.0752&	0.2106&	0.0945\\

GT (CORONA)& 
         \textbf{0.0628}&	\textbf{0.0318}&	\textbf{0.0931}&	\textbf{0.0294}&	\textbf{0.1468}&	\textbf{0.0471}&
         \textbf{0.3072}&	\textbf{0.1528}&	\textbf{0.4165}&	\textbf{0.1988}&	\textbf{0.5504}&	\textbf{0.2619}&
         \textbf{0.1203}&	\textbf{0.1028}&	\textbf{0.1928}&	\textbf{0.1202}&	\textbf{0.3011}&	\textbf{0.1221}\\
        
         \Xhline{0.8pt}
    \end{tabular}
\label{different_combination}
\end{table*}
\end{footnotesize}

\renewcommand{\arraystretch}{1.7}
\setlength{\tabcolsep}{1.3pt}
\begin{footnotesize}
\begin{table*}[ht]
\caption{Ablation study of different components on three datasets in terms of \textit{Recall@}10/20/50, and \textit{NDCG@}10/20/50.}
\vspace{-0.4cm}
\centering
    \begin{tabular}{@{} c|>{\columncolor{gray!30}}cc>{\columncolor{gray!30}}cc>{\columncolor{gray!30}}cc|>{\columncolor{gray!30}}cc>{\columncolor{gray!30}}cc>{\columncolor{gray!30}}cc|>{\columncolor{gray!30}}cc>{\columncolor{gray!30}}cc>{\columncolor{gray!30}}cc @{} }
    \Xhline{0.8pt}
        Datasets & \multicolumn{6}{c|}{Netflix} &  \multicolumn{6}{c|}{MovieLens} & \multicolumn{6}{c}{Amazon-book}	  \\ 
        \hline
        Methods  & R@10 & N@10 & R@20 & N@20 & R@50 & N@50 & R@10 & N@10 & R@20 & N@20 & R@50 & N@50 & R@10 & N@10 & R@20 & N@20 & R@50 & N@50 \\ \hline
w/o Preference-assisted Retrieval&
0.0576& 0.0227& 0.0848& 0.0343& 0.1295& 0.0431&
0.2886& 0.1375& 0.3873& 0.1970& 0.5306& 0.1829&
0.1112& 0.0747& 0.1535& 0.0909& 0.2228& 0.1036\\
w/o Intent-assisted Retrieval&
0.0529& 0.0208& 0.0785& 0.0329& 0.1006& 0.0418&
0.2694& 0.1092& 0.3590& 0.1913& 0.5129& 0.1768&
0.1037& 0.0764& 0.1643& 0.0937& 0.2028& 0.0961\\
w/o GNN-enhanced Retrieval&
\underline{0.0591}& \underline{0.0264}& \underline{0.0917}& \underline{0.0359}& \underline{0.1304}& \underline{0.0457}&
\underline{0.2958}& \underline{0.1434}& \underline{0.4088}& \underline{0.1943}& \underline{0.5531}& \underline{0.1905}&
\underline{0.1189}& \underline{0.0814}& \underline{0.1729}& \underline{0.0968}& \underline{0.2311}& \underline{0.1106}\\
w/o Preference Reasoning&
0.0506& 0.0187& 0.0759& 0.0297& 0.0954& 0.0386&
0.2532& 0.1158& 0.3316& 0.1741& 0.4819& 0.1703&
0.0933& 0.0653& 0.1490& 0.0875& 0.1986& 0.0887\\
w/o Intent Reasoning&
0.0451& 0.0142& 0.0735& 0.0264& 0.0933& 0.0371&
0.2317& 0.1173& 0.3221& 0.1476& 0.4609& 0.1659&
0.0789& 0.0621& 0.1215& 0.0839& 0.1982& 0.0835\\
Full CORONA&
\textbf{0.0616}& \textbf{0.0279}&	\textbf{0.0938}&	\textbf{0.0416}&	\textbf{0.1452}& \textbf{0.0487}&
\textbf{0.3033}&	\textbf{0.1565}&	\textbf{0.4214}&	\textbf{0.2017}&	\textbf{0.5745}&	\textbf{0.2507}&
\textbf{0.1206}&	\textbf{0.0855}&	\textbf{0.1857}& \textbf{0.1089}&	\textbf{0.3048}&	\textbf{0.1299}
\\

\Xhline{0.8pt}
    \end{tabular}
\label{ablation_study}
\end{table*}
\end{footnotesize}

\subsection{Hyperparameter Analysis (RQ4)}
We investigate the impact of key hyperparameters in this subsection, including the size of retrieved subgraph \(k\), the temperature parameter of LLM $\tau$, the dimension of distance encoding $\dim(e)$ and the hidden dimension of GNN. For each dataset, we use R@20 as the evaluation metric, and the results are shown in Figure~\ref{k_experiments}. 

\subsubsection{Size of Retrieved Subgraph \(k\)}
We vary the value of \(k\) from $1,000$ to $4,000$, and observe that the performance improves as \(k\) increases up to a certain point, after which it begins to decline. For all datasets and metrics the best performance is achieved when $k$ is around $3,000$. 
% As shown in Figure~\ref{k_experiments}, smaller \(k\) values fail to capture enough relevant information, while larger \(k\) values introduce noise.
% Setting \(k\) to $3,000$ strikes the best balance between capturing relevant data and minimizing noise.

\subsubsection{Temperature $\tau$ of LLM}
\begin{figure}
    \centering
    \includegraphics[width=\linewidth]{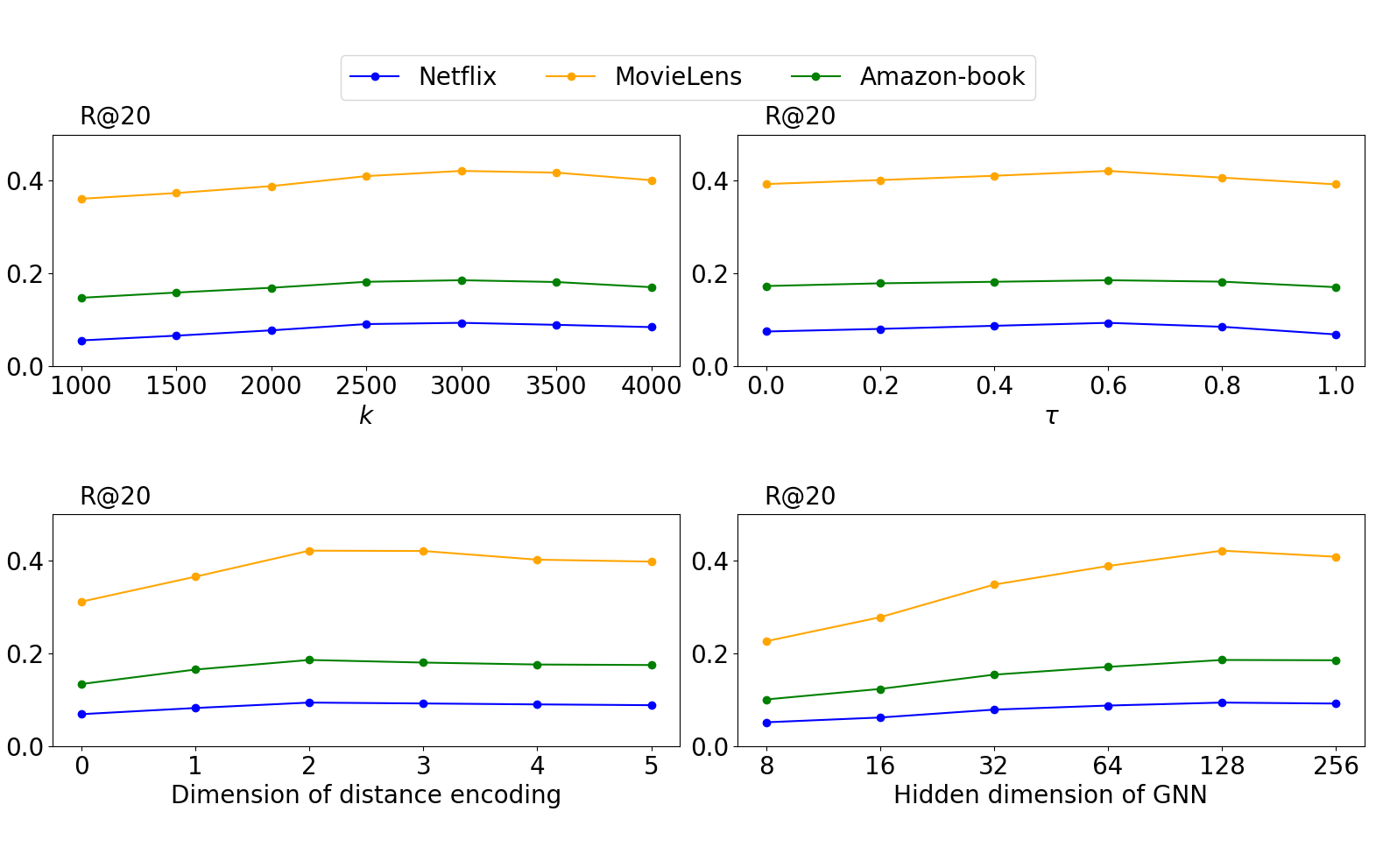}
    \vspace{-1cm}
    \caption{Hyperparameter experiments on three datasets in terms of \textit{Recall@}20.}
    \label{k_experiments}
% \vspace{-0.7cm}
\end{figure}

Follow existing work~\cite{wei2024llmrec}, we conduct experiments on the temperature parameter $\tau$, which controls the randomness of text generation. Higher values lead to greater diversity and creativity, while lower values produce more deterministic outputs. We experiment with $\tau$ values of \{0, 0.2, 0.4, 0.6, 0.8, 1\}. As illustrated in Figure~\ref{k_experiments}, increasing $\tau$ slightly enhances most metrics at the beginning, but further increases result in a decline.

\subsubsection{Dimension of Distance Encoding $\dim(e)$}
We evaluate the impact of different dimensions of distance encoding on model performance, which plays a crucial role in the subgraph retrieval process. A smaller distance encoding dimension weakens the model’s ability to differentiate neighbors at varying distances, while a larger dimension may lead to overfitting. We test distance encoding dimensions ranging from 0 to 5, and the results align with expectations, with the best performance achieved at a dimension of 2.

\subsubsection{Hidden Dimension of GNN}
We test hidden dimensions of \{8, 16, 32, 64, 128, 256\} on the message-passing component in GNN-enhanced retrieval. The results show improved performance as the hidden dimension increases, peaking at 128 dimensions, after which a slight decline is observed.

\subsection{Efficiency Analysis (RQ5)}
\begin{figure}
    \centering
    \includegraphics[width=\linewidth]{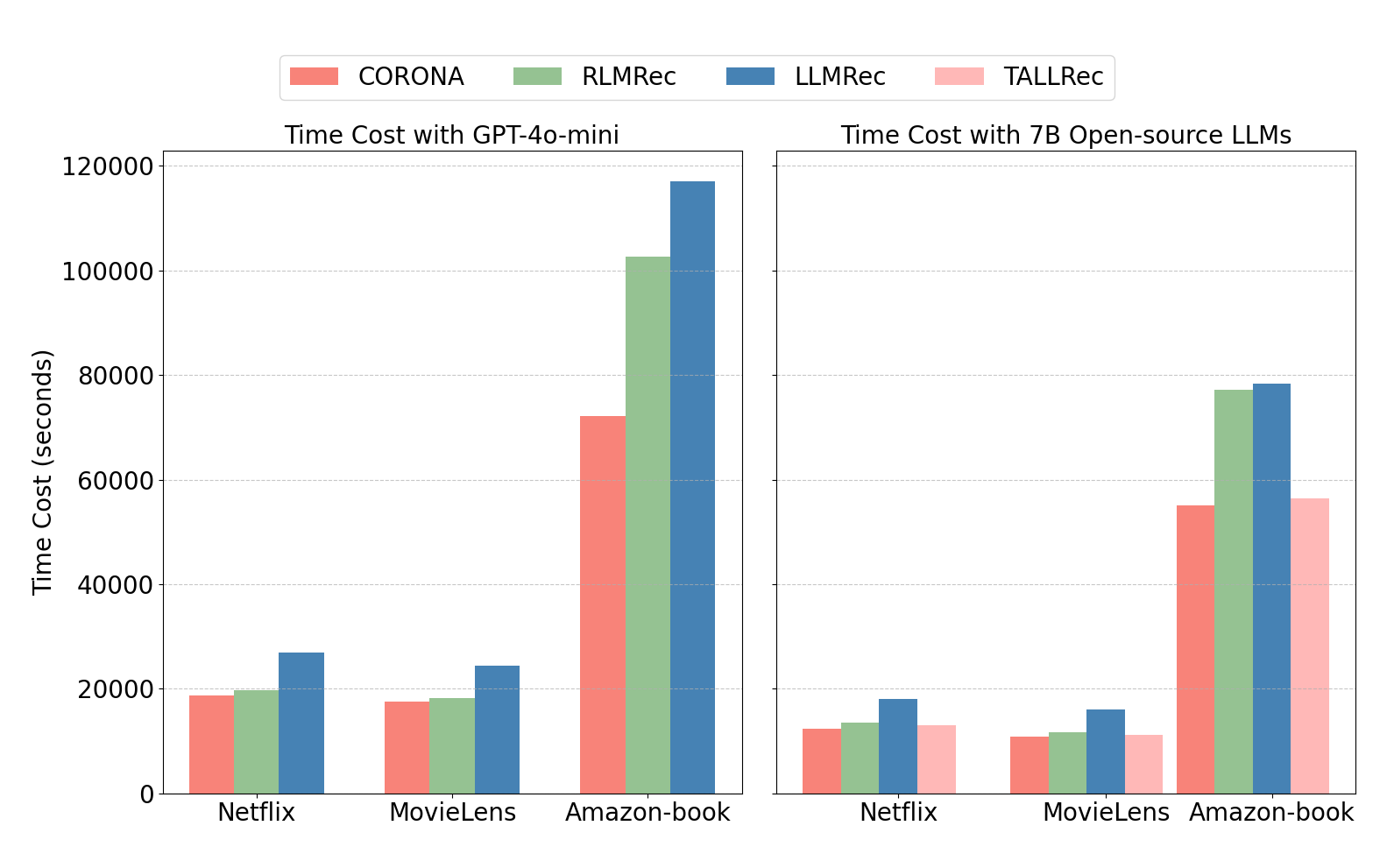}
    \vspace{-1cm}
    \caption{Total time cost (in seconds) of different LLM-based recommendation methods.} 
    \label{efficiency}
\vspace{-0.6cm}
\end{figure}

To answer RQ5 and validate the efficiency of CORONA, we measure the total running time on three datasets, and compare it with three LLM-based methods. Since TALLRec needs to fine-tune LLMs, we only compare it under open-source LLMs. The results are represented in Figure~\ref{efficiency}. From the results, we can observe that: (1) Although CORONA involves two LLM inference steps, its inference focuses only on users, avoiding the large-scale inference over items or interactions, which ensures better efficiency compared to other methods. (2) CORONA is more efficient with open-source LLMs than GPT-4o-mini that suffers from additional overhead from network transmission. (3) CORONA with GPT-4o-mini consumes approximately 1 US cent and less than 1.5 seconds per user for inference. With locally deployed LLMs, the inference time can be further reduced to less than one second. While lightweight open-source LLMs show a slight performance drop, we interpret this as a balance between performance and efficiency, allowing users to choose the version that best fits their needs. (3) LLMRec needs to augment over all items, users and interactions in the dataset, and thus takes longer processing time, especially for the Amazon-book dataset with denser interactions. 
% \begin{table}[h]
% \centering
% \caption{Time Cost Comparison Across Datasets and Methods}
% \resizebox{\columnwidth}{!}{%
% \begin{tabular}{@{}lccc@{}}
% \toprule
% \textbf{Methods} & \textbf{Netflix (s)} & \textbf{MovieLens (s)} & \textbf{Amazon-book (s)} \\ \midrule
% CORONA with GPT   & 18.725k & 17.500k & 72.121k \\
% CORONA with Vicuna & 12.318k & 10.879k & 55.081k \\
% LLMRec with GPT    & 26.847k & 24.465k & 117.024k \\
% LLMRec with Vicuna & 18.068k & 16.089k & 77.217k \\
% RLMRec with GPT    & 19.676k & 18.132k & 102.596k \\ 
% RLMRec with Vicuna & 13.464k & 11.652k & 78.368k \\
% \bottomrule
% \end{tabular}%
% }
% \label{tab:time_cost_comparison}
% \end{table}

\subsection{Case Study (RQ6)}
% \begin{figure*}[ht]
%     \setlength{\abovecaptionskip}{-0.1cm} 
%     \centering
%     \begin{subfigure}[b]{0.4\linewidth}
%         \centering
%         \includegraphics[width=\linewidth]{Figs/case_a.pdf}
%         \vspace{-0.7cm}
%         \caption{Case of a male user.}
%         \label{subfig_case_a}
%     \end{subfigure}
%     \hfill % 用于在两个子图之间加入间隙
%     \begin{subfigure}[b]{0.4\linewidth}
%         \centering
%         % \vspace{0.5cm}
%         \includegraphics[width=\linewidth]{Figs/case_b.pdf}
%         \vspace{-0.7cm}
%         \caption{Case of a female user.}
%         \label{subfig_case_b}
%     \end{subfigure}
%     \caption{Case study of recommendation with our proposed CORONA. Here we mark the correct items in the final recommendation list in green, and the key information related to the correct items are highlighted in red. }
%     \label{case-study}
% \end{figure*}

\begin{figure}[ht]
    \centering
    \begin{subfigure}[b]{\linewidth}
        \centering
        \includegraphics[width=1.023\linewidth]{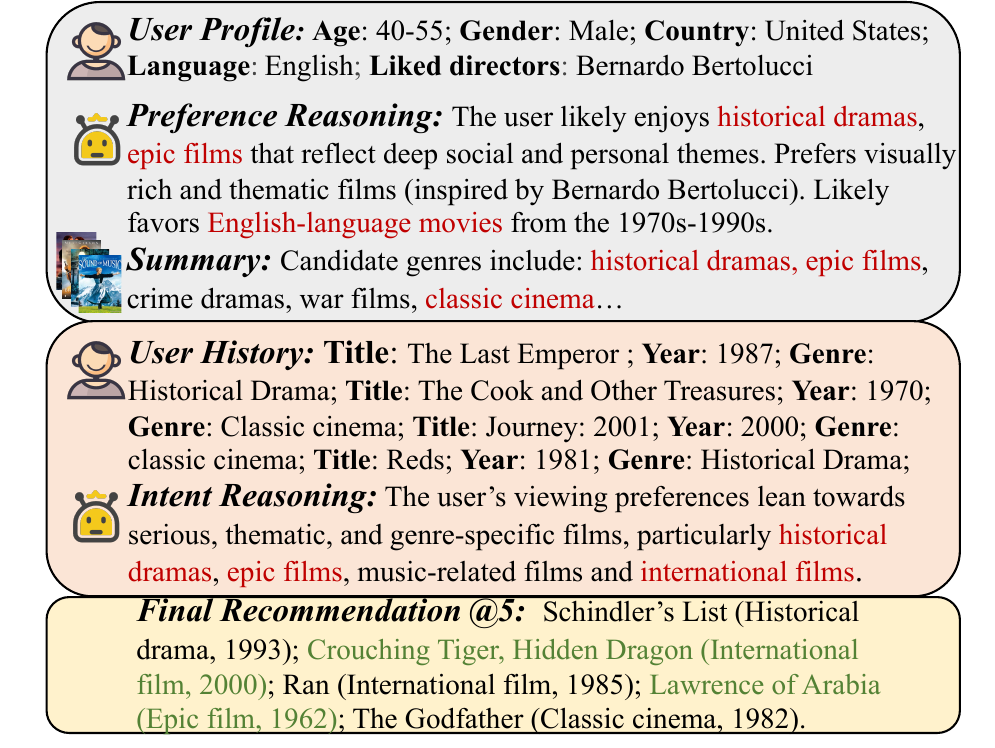}
        \caption{Case of a male user.}
        \label{subfig_case_a}
    \end{subfigure}
    \vspace{-0.2cm} % 子图间距
    \begin{subfigure}[b]{\linewidth}
        \centering
        \includegraphics[width=1.023\linewidth]{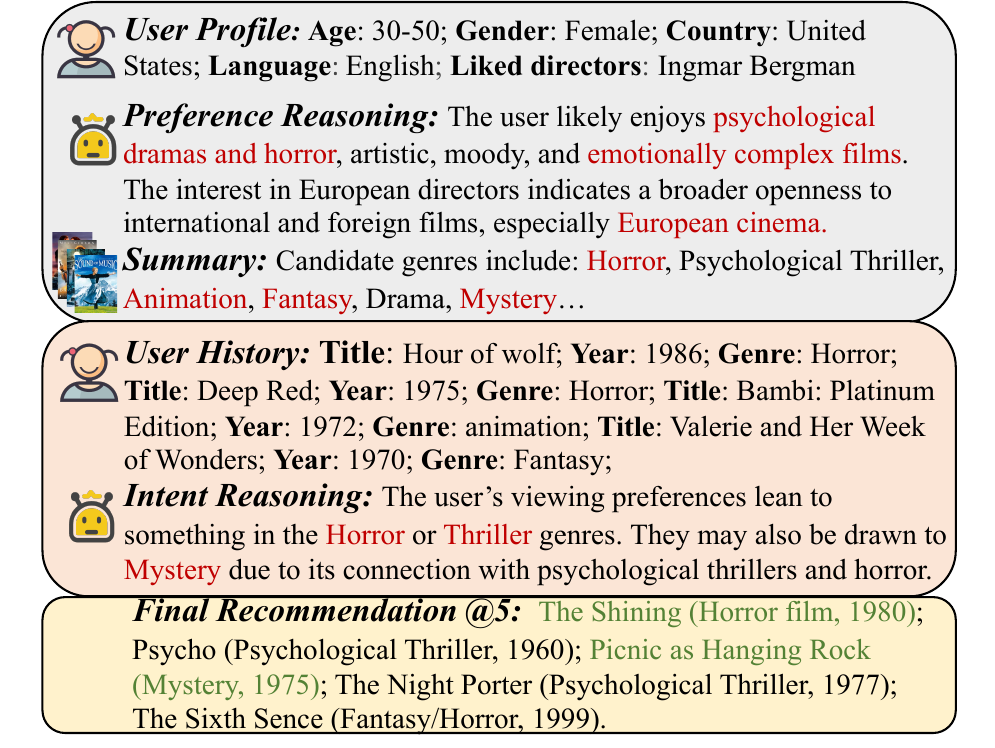}
        \caption{Case of a female user.}
        \label{subfig_case_b}
    \end{subfigure}
    \caption{Case study of recommendation with our proposed CORONA. Here we mark the correct items in the final recommendation list in green, and the key information related to the correct items are highlighted in red.}
    \label{case-study}
\end{figure}

To illustrate the inference process of our proposed \modelname framework in an understandable natural language form and further verify whether the LLM-based reasoner can generate plausible user preferences and intents, we present two examples from the test set of Netflix dataset in Figure~\ref{case-study}. In the examples, we show the user profile, the preference reasoning, item summary of preference-assisted retrieval, interaction history, intent reasoning, and the final recommendation. The results show that the LLM-based reasoners can effectively infer relevant content leveraging their commonsense knowledge and enhance recommendation performance.

\section{Conclusion}
In this paper, we present a novel paradigm to incorporate LLMs into recommendation systems, where LLMs are employed for coarse-grained reasoning to assist in retrieval across the entire item set. The proposed CORONA framework has a carefully designed three-stage retrieval process that progressively refines the selection at different levels of granularity. CORONA harnesses the reasoning power of LLMs for preference and intent inference, combined with GNNs for efficient recommendations. Extensive experiments confirm the effectiveness of the CORONA framework and validate its design. Future work may extend our framework to larger-scale industrial scenarios with more stages of retrieval. Additionally, exploring different LLMs to find the most cost-effective implementation could be valuable. It is also possible to finetune an open-source LLM for more accurate preference or intent reasoning.%, such as using recommendation-specific language models instead of GPT-4-o-mini as the reasoner to reduce the overall cost.

\section*{Acknowledgments}
This work was supported by the National Key Research and Development Program of China (No.2023YFC3303800).

\clearpage
\bibliographystyle{ACM-Reference-Format}
\balance
\bibliography{citation}

\clearpage
\appendix

\end{document}